\documentclass[%
 reprint,
superscriptaddress,
 amsmath,amssymb,
prb,
floatfix,
]{revtex4-2}
\usepackage[final]{hyperref}
\hypersetup{
	colorlinks=true,       % false: boxed links; true: colored links
	linkcolor=blue,        % color of internal links
	citecolor=magenta,     % color of links to bibliography
	filecolor=magenta,     % color of file links
	urlcolor=blue         
}
\usepackage{booktabs}
\usepackage{dsfont}
\usepackage{mathtools}
\usepackage{float}
\usepackage{subfigure}
\usepackage{braket}
\usepackage{graphicx}% Include figure files
\usepackage{dcolumn}% Align table columns on decimal point
\usepackage{bm}% bold math

\usepackage{natbib}
\begin{document}

\preprint{APS/123-QED}

\title{Hyperfine Interaction in a MoS$_2$ Quantum Dot: Decoherence of a Spin-Valley Qubit}% Force line breaks with \\

\author{Mehdi Arfaoui}
 \altaffiliation[]{\href{mailto:arfaouimehdi20@yahoo.fr}{arfaouimehdi20@yahoo.fr}}%Lines break automatically or can be forced with \\
 \affiliation{Condensed Matter Physics Laboratory,
Department of Physics, Faculty of Sciences of Tunis, University 
Tunis El Manar, University Campus, 1060 Tunis, Tunisia.}
\author{Sihem Jaziri}%
 \email{sihem.jaziri@fsb.rnu.tn}
\affiliation{%
Condensed Matter Physics Laboratory,
Department of Physics, Faculty of Sciences of Tunis, University 
Tunis El Manar, University Campus, 1060 Tunis, Tunisia.
}%
\affiliation{%
Materials Physics Laboratory, Faculty of Sciences of Bizerte, University of Carthage, 7021 Zarzouna, Tunisia.
}%
\break

\date{\today}% It is always \today, today,

\begin{abstract}
A successful and promising device for the physical implementation of electron spin-valley based qubits is the Transition Metal Dichalcogenide monolayer (TMD-ML) semiconductor quantum dot. The electron spin in TMD-ML semiconductor quantum dots can be isolated and controlled with high accuracy, but it still suffers from decoherence due to the unavoidable coupling with the surrounding environment, such as nuclear spin environments. A common tool to investigate systems like the one considered in this work is the density matrix formalism by presenting an exact master equation for a central spin (spin-qubit) system in a time-dependent and coupled to a nuclear spin bath in terms of hyperfine interaction. The master equation provides a unified description of the dynamics of the central spin. Analyzing this in more detail, we calculate fidelity loss due to the Overhauser field from hyperfine interaction in a wide range of number of nuclear spin $\mathcal{N}$.
\end{abstract}

\maketitle

%\tableofcontents

\section{Introduction}

Modern life is powered by information, which leads to an increase in the demand for computational power so much that new generations of semiconductor technologies are employed consistently. Since then, there have been some reports headed for the development and improvement of valleytronic devices \cite{lee2017valley, Valleytronics}, and TMD qubits, including valley qubits, spin qubits, spin-valley qubits and even impurity based qubits \cite{Korm_nyos_2014, Kormanyos_2018, Impurity_2018, PhysRevB, pearce2017electron, pawlowski2018valley, Brooks_2017}. Quantum dots in a monolayer transition metal dichalcogenide (TMD-QD) such as Molybdenum Disulfide MoS$_{2}$ is a new pattern for qubit and holds the promises for new quantum devices but it still suffer from decoherence. Precisely, the major obstacles that are addressed pose a challenge to researchers, noise hinders the transmission of quantum signals. One of the essential tools in the study of spin-valley qubits in quantum dots is stability, which is delicate and very sensitive to the disturbances of a noisy environment. Indeed, the inevitable coupling and that each tiny interaction of these objects with their environment very quickly destroys the phase relations (superposition of incompatible states) between the quantum states until they become classical states. The electron spin in a quantum dot has two main decoherence channels, a (Markovian) phonon-assisted relaxation channel, due to the presence of spin-orbit interaction, and a (non-Markovian) spin bath constituted by the spins of the nuclei in the quantum dot that interact with the electron spin via the hyperfine interaction \cite{coish2004hyperfine, fuchs2012spin,wu2016spin,RevModPhys.85.79} where the number of nuclear spins ranges from $\sim$ $10^2$ up to $10^6$ and full polarized baths are employed to facilitate qubit operations and extend coherence times.\\
A powerful tool for dealing with such systems is provided by an alternative way of deriving an exact master equation is the second-order time-convolutionless (TCL) master equation \cite{PhysRevA.81.042103,Shenarticle2014,smirne2010nakajima} of initial bath states, including the full polarized bath. Although we will focus on the example of spin qubits, our results are potentially applicable to any central spin problem.\\
In this work, we solve the central spin problem using the TCL master equation. This equation enables us to study the dynamics of the central spin, and more interestingly showing the fidelity of the spin qubit to investigate the non-Markovian character in the dynamical decoherence of open quantum systems. Motivated by this consideration, in this paper, we consider a fluctated environment with a full polarized nuclear spins to study a decoherence effect at the qubit system which can be solved exactly.\\
The paper is organized as follows. In Sec.\eqref{Qubit_candidat} we discuss the best candidate monolayer for a spin qubit. To obtain realistic values of the parameters appearing in the theory we have performed density functional theory (DFT) calculations. Next, the TMD-QD Hamiltonian is given with an external magnetic field perpendicular to the quantum dot is considered, and numerical solutions to the necessary external field strengths at a given quantum dot radius are shown at which a spin-degenerate state within a given valley is expected. In Sec.\eqref{ME} we derive an effective Hamiltonian describing this single electron spin and $\mathcal{N}$ spin-nuclei in interaction.  Our approach is based on deriving an appropriate exact non-Markovian time-convolutionless (TCL) master equation (ME) describing the evolution of the qubit system. We further apply this result to quantify the ﬁdelity loss that the noise induces, and Sec.\eqref{conclusion} concludes with a discussion about our finding.
\section{\label{Qubit_candidat}Single Electron Quantum Dots as Spin Qubits}
\begin{figure}
\centering
\subfigure[]{\includegraphics[width=0.55\linewidth]{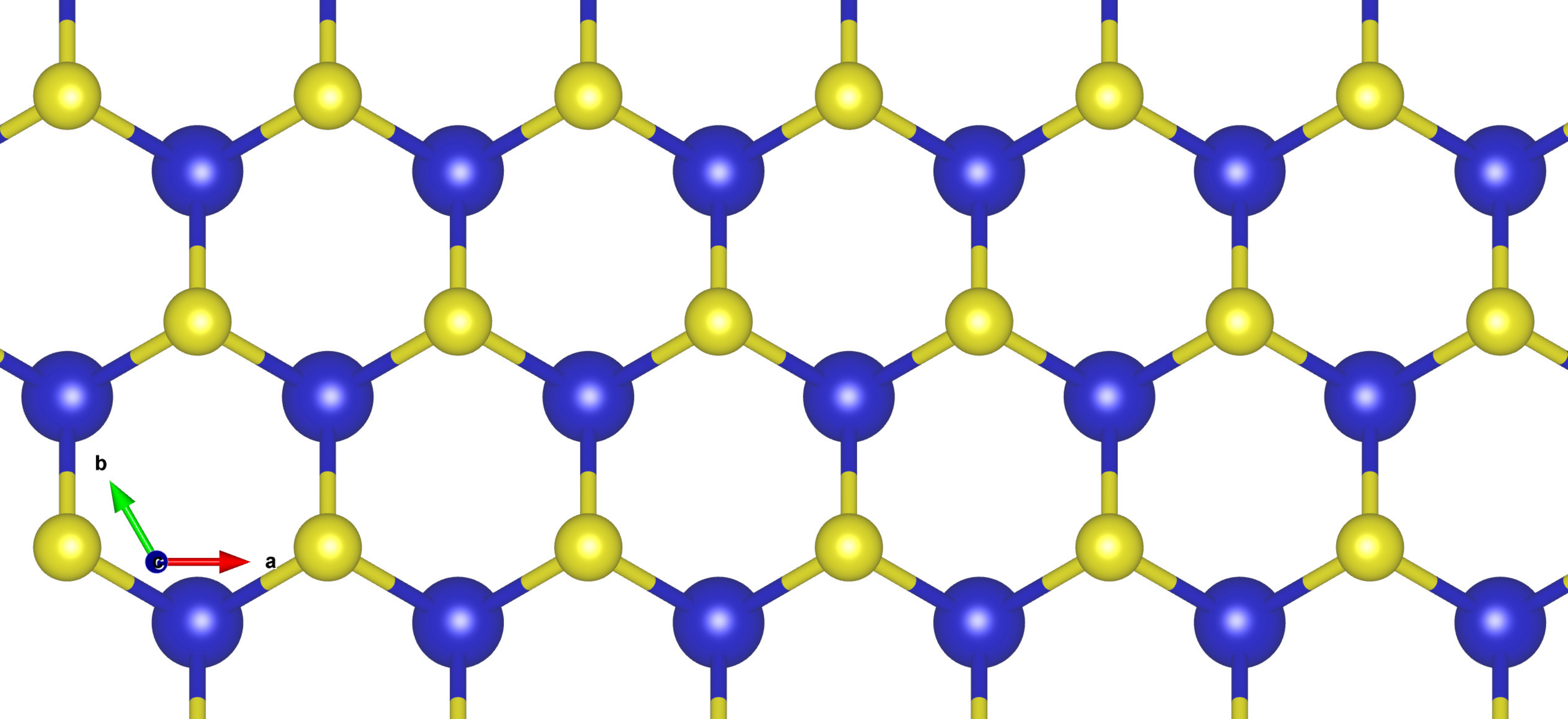}}
\hspace{0.12\linewidth}
\subfigure[]{\includegraphics[width=0.3\linewidth]{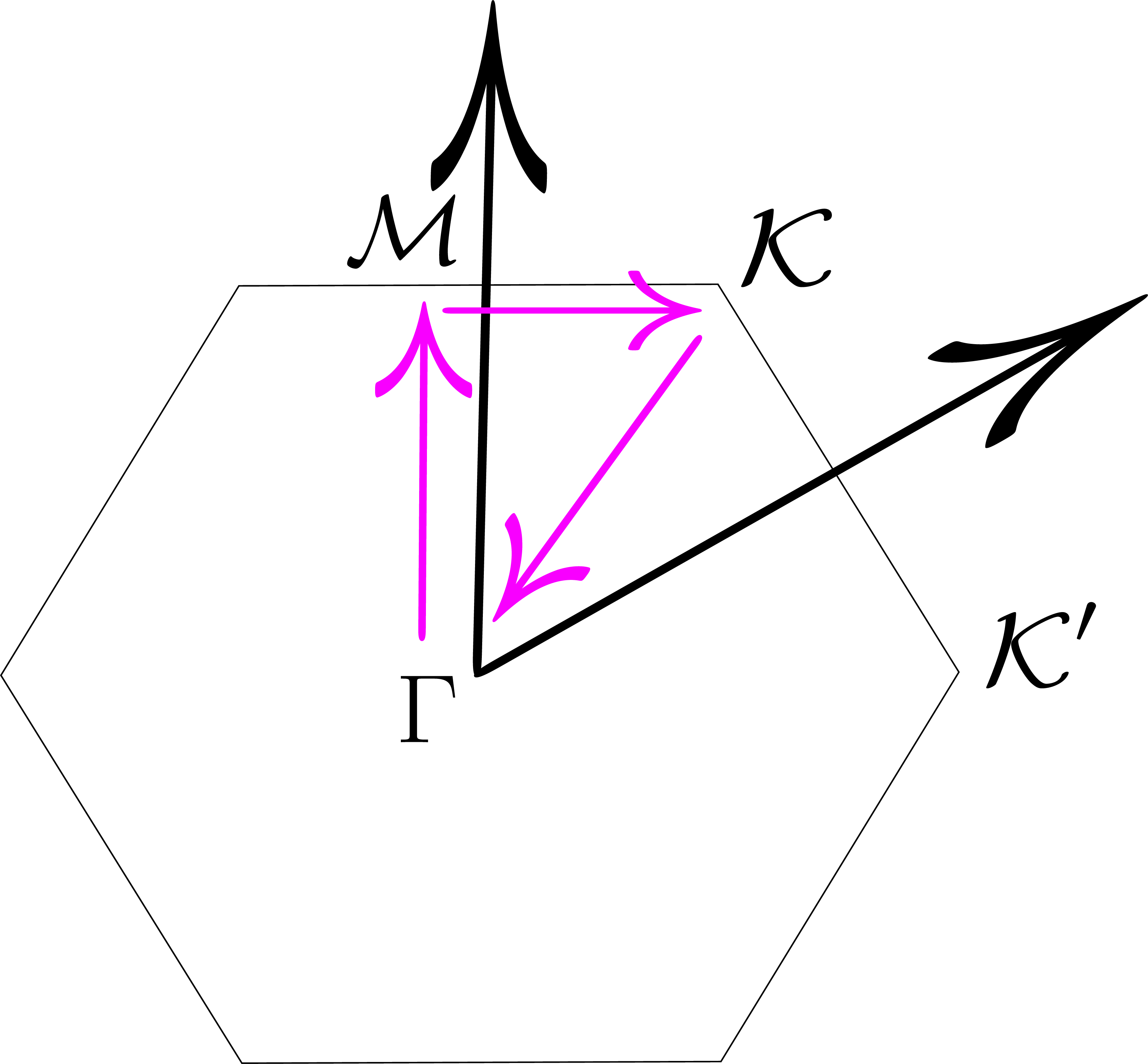}}
\subfigure[]{\includegraphics[width=\linewidth]{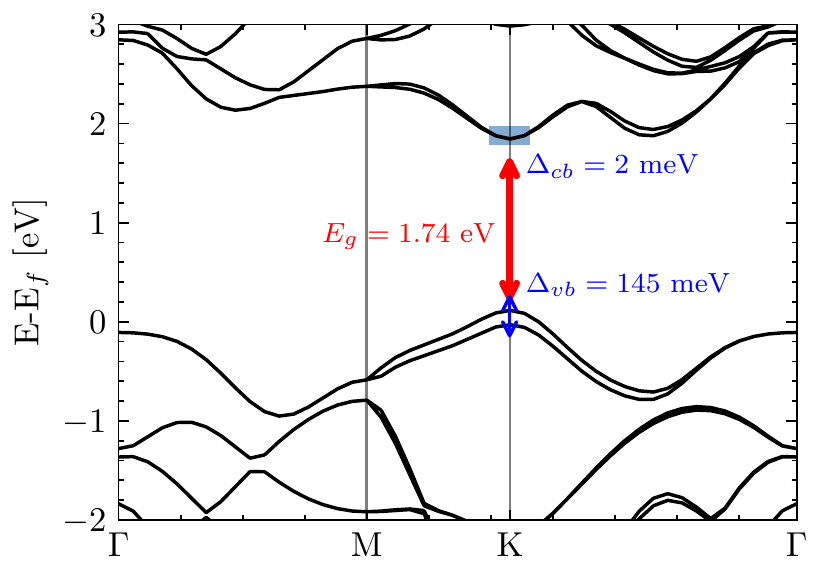}\label{bs}}
\subfigure[]{\includegraphics[width=\linewidth]{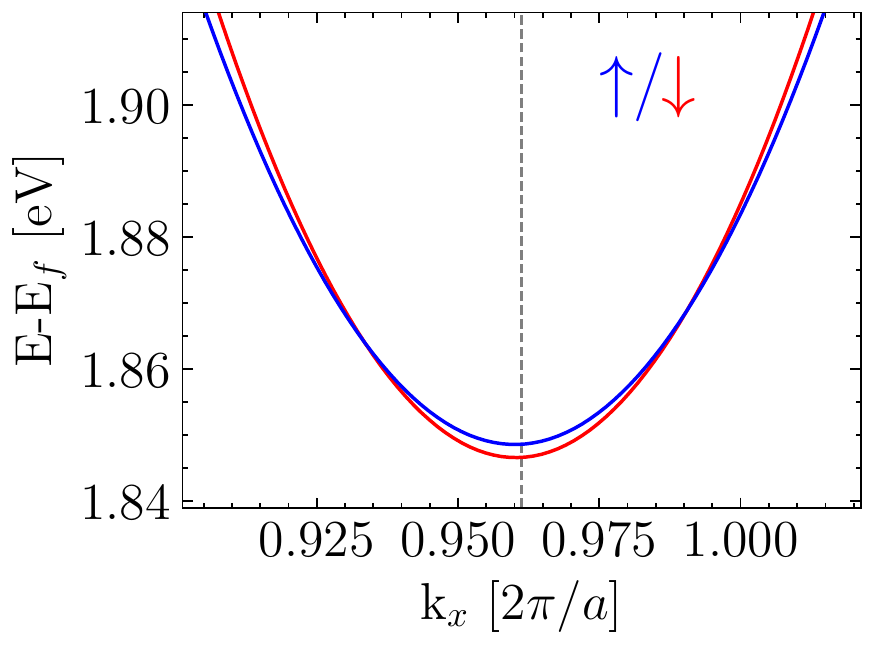}\label{cb_bs}}
\caption{(a) Top view of ML MoS$_2$. Blue and yellow balls represent Mo and S atoms, respectively. (b) Brillouin zone with high symmetry k-path. (c) Band structure of an ML-MoS$_2$ obtained from DFT calculations. Showing both the CB  and VB for both spin up and spin down states. Here the band gap $E_g$ = 1.74 eV and intrinsic SOC splitting at the $\mathcal{K}$ valley for both the CB ($\Delta_{cb}$ = 2 meV) and VB ($\Delta_{vb}$ = 145 meV) are marked. We zoom in the blue rectangle where is shown in the Fig.\eqref{cb_bs}. (d) Spin resolved conduction band (red: $|0\rangle=\left|\mathcal{K}^{\prime} \uparrow\right\rangle$ and $|1\rangle_{-}=|\mathcal{K} \downarrow\rangle$, blue: $|0\rangle_{-}=|\mathcal{K} \uparrow\rangle$ and $\left.|1\rangle=\left|\mathcal{K}^{\prime} \downarrow\right\rangle\right)$ around the $\mathcal{K}$ valley in the $\mathrm{BZ}$ demonstrating the spin crossings present in MoS$_{2}$, the $\mathcal{K}^{\prime}$ valley may be visualized simply by the time-reversal of the given band structure. a is the lattice parameter.}
\end{figure}

A huge effort is underway to develop semiconductor nanostructures as low noise hosts for qubits. To provide the possibility for Transition Metal Dichalcogenide (TMD) monolayer candidates to use as a host for a spin qubit, the two spin states ($\ket{\uparrow}, \ket{\downarrow}$) selected for the desired qubit need to be degenerate, or tuneable by some external influence (Magnetic field) into a degeneracy. Understanding the band structure and external field replies is a requirement for achieving qubits with carriers in TMD ML. However, the strong intrinsic spin–orbit coupling (SOC) within this material (TMD)  exhibits the opposite influence and makes this a non-trivial task. Indeed, the large splitting occurred of the spin states, $\sim meV$, within the same valley ($\mathcal{K}(\mathcal{K}^{\prime}$)) in the conduction band (CB),  means that a single electron within a quantum dot (QD) in TMD will not naturally explain the required degeneracies wanted for a spin qubit. Favorably, the band crossing seen in the spin-resolved CB structures in MoS$_2$-ML which submit that it is reasonable to accomplish spin degeneracy localized within a given valley ($\mathcal{K}(\mathcal{K}^{\prime}$)), see Fig. \eqref{cb_bs}. Consequently, Such spin-degenerate regimes allow the possibility of realizing the desired spin qubits in the ML-MoS$_2$ quantum dot \cite{Korm_nyos_2014}.

\subsection{Modeling the Quantum Dot: Theoretical Model}\label{Qubit_selection}
The eigenenergys of a single electron confined in a TMD quantum dot by the parabolic potential in a perpendicular magnetic field \textbf{B} = (0, 0, $B_{z}$), $ B_{z} > 0,$ at the $\mathcal{K}$ or $\mathcal{K}^{\prime}$ valleys may be obtained by solving the effective low energy Hamiltonian\cite{Korm_nyos_2014}
\begin{align}\label{parabolic_mag}
\mathcal{H}_{B_{z}}^{\tau,s}=&
\frac{\hbar^{2} q_{+}q_{-}}{2m_{eff}^{\tau,s}}+ \frac{1}{2}m_{eff}^{\tau,s} \Omega_{\tau,s}^{2} \,  r^{2}-\frac{1}{2}w_c^{\tau,s}\,\ell_z
+\tau s {\Delta_{cb} \over 2}\\
&+{1 \over 2} \tau g_{vl}\, \mu_{B} B_{z}
+{1 \over 2} s g_{sp}\, \mu_{B} B_{z}\notag
\end{align}
with $m_{eff}^{\tau,s}$ is the effective mass of the conduction band, where it's shown in Table \eqref{parameter_dft} (note that $m_{eff}^{\tau,-s}=m_{eff}^{-\tau,s}$) which obtained from fitting the DFT band structure at high
symmetry $\mathcal{K}$ point of the first Brillouin zone (BZ), see Fig. \eqref{bs}. The band structure calculations were performed by the DFT and the Full Potential–Linearized Augmented Plane Wave (FP-LAPW) method using Wien2k code~\cite{BLAHA1990399, blaha2001wien2k}, and the generalized gradient approximation (GGA) framework with a Perdew-Burke-Ernzerhof (PBE) functional is used for the exchange correlation potential~\cite{perdew1992atoms}.
\begin{table*}
\caption{\label{parameter_dft} Effective masses, CB spin-splitting, VB spin-splitting, the energy of the band gap, valley and spin g-factor for ML-MoS$_2$ appearing in Hamiltonian \eqref{parabolic_mag}. m$_e$ is the free-electron mass.}
\begin{ruledtabular}
\begin{tabular}{cccccccc}
 & $\Delta_{cb}$ {[}meV{]} & $\Delta_{vb}$ {[}meV{]} & m$_{eff}^{\mathcal{K},\uparrow}$/m$_{e}$ & m$_{eff}^{\mathcal{K},\downarrow}$/m$_{e}$ & E$_g$[eV]& $g_{vl}$&$g_{sp}$\\ \hline
MoS$_2$ & 2& 145 & 0.54& 0.49 & 1.74&0.75\cite{Korm_nyos_2014}&1.98\cite{Korm_nyos_2014} \\
\end{tabular}
\end{ruledtabular}
\end{table*}

$\tau$ and s denote the index, which takes the value 1 (-1) indicated by the valley $\mathcal{K}$( $\mathcal{K^{'}}$) and the spin $\ket{\uparrow}(\ket{\downarrow})$, respectively.  $q_{\pm}=q_{x} \pm i q_{y}$ is the wave number operators, where $q_{k}=-i \partial_{k}$, $g_{vl}$ and  $g_{sp}$ is the valley and spin g-factor, $\ell_z$ is the z component of the orbital moment, $\mu_{B}$ is
Bohr's magneton and $\Omega_{\tau, s}$ is the effective frequency, giving by,
\begin{equation}
\Omega_{\tau, s} = \sqrt{(\omega_0^{\tau, s})^2 + \frac{(\omega_c^{\tau, s})^2}{4}}
\end{equation}
where, $\omega_0^{\tau, s} = \hbar /(m_{eff}^{\tau, s}R^2)$ (R the QD radius), and $\omega_c^{\tau, s}=(e B_{z})/(m_{eff}^{\tau, s})$ denote respectively the parabolic confinement and the cyclotron frequency.

Thus, the QD levels as a function of out-of-plane magnetic field $B_z$ and QD radius R are given as
\begin{widetext}
\begin{equation}
\mathcal{E}_{n,\ell}^{{\tau,s}} = \hbar\, \Omega_{\tau,s} (2n + 1+ \mid \ell \mid)-\frac{1}{2} \hbar w_c^{\tau,s} \ell  +
\tau s {\Delta_{cb} \over 2}
+{1 \over 2} \tau g_{vl}\, \mu_{B} B_{z} +
{1 \over 2} s g_{sp}\, \mu_{B} B_{z}
\end{equation}
\end{widetext}
where n = 0, 1, \ldots is the radial quantum number; $\ell$= -n, -n + 2, \ldots, n - 2, n is the angular momentum quantum number.\\
The Hamiltonian wavefunction \eqref{parabolic_mag} can be written as follows
\begin{align}
\psi^{\tau,s}_{n,\ell}&(r,\phi)
= A_{n,\mid\ell\mid}{exp\,(i\ell \phi)\over \sqrt{2\pi}} \left({m_{eff}^{\tau,s}\Omega_{\tau,s}\over 2\hbar}r^{2}\right)^{\mid{\ell/ 2}\mid}\,\\ &\times exp\left( -{m_{eff}^{\tau,s}\Omega_{\tau,s}\over 4\hbar}r^{2}\right)
L^{\mid\ell\mid}_{n}\left({m_{eff}^{\tau,s}\Omega_{\tau,s}\over 2\hbar}r^{2}\right)\chi_s(\sigma_z)\notag
\end{align}
with $A_{n,\mid\ell\mid}$ is the normalization coefficient and $L^{\mid\ell\mid}_{n}\left(x\right)= {1 \over \ell !} x^{-\mid\ell\mid}e^{x}{d^{n}\over d x^{n}}\left( x^{n+\mid\ell\mid}e^{-x}\right)$ is the associated Laguerre polynomials. $\chi_s(\sigma_z)$ is the eigenstate of the spin operator $S_z = \hbar\sigma_z/2$, where $\sigma_z = \begin{pmatrix}
	1 & 0 \\
	0 & -1 
\end{pmatrix}$ is the Pauli matrice.

\begin{figure}
\centering
\includegraphics[width=\linewidth]{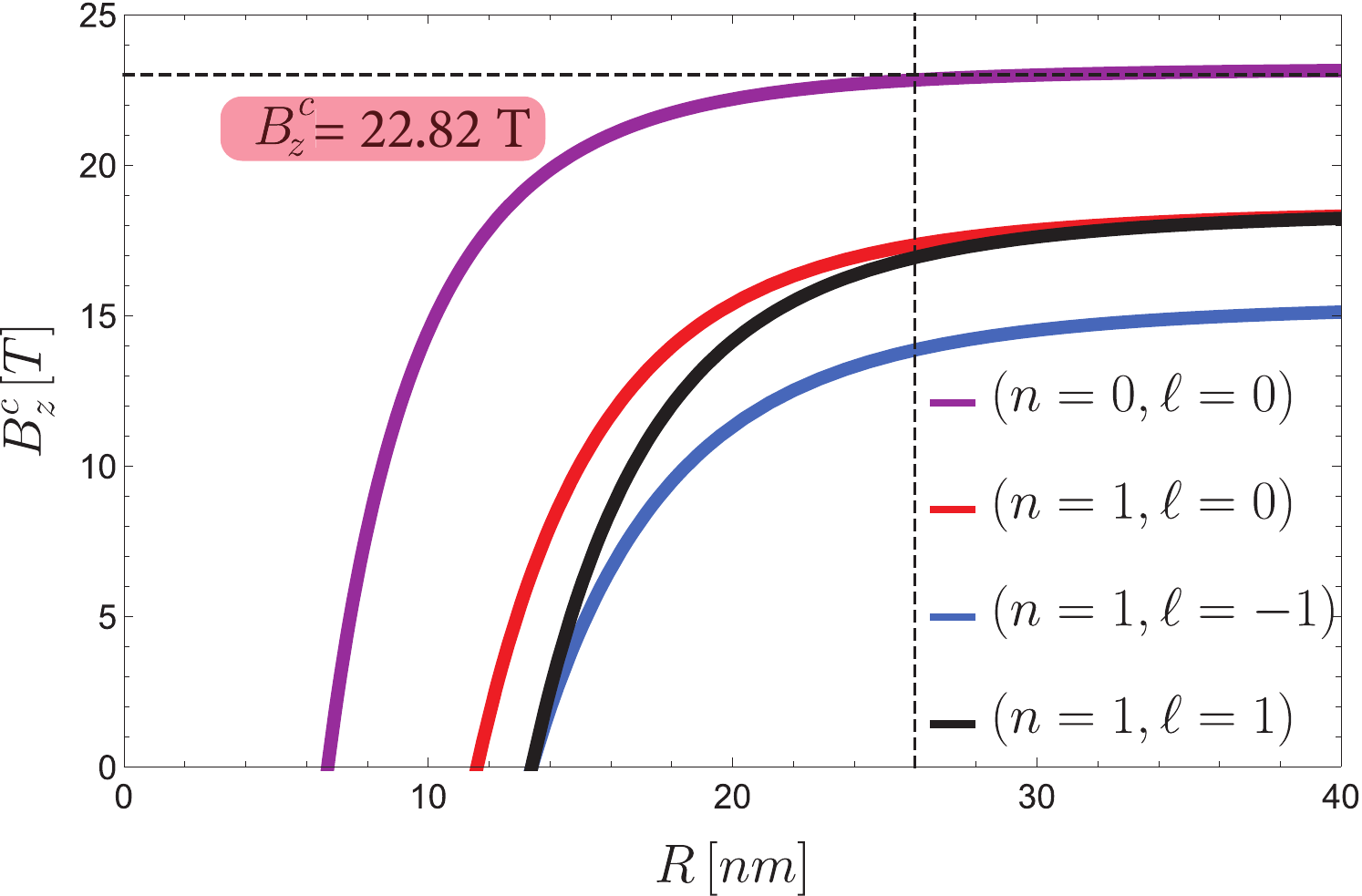}
\caption{Spin-degeneracy curves of critical out-of-plane magnetic
field $B_z^{c}$ as function to QD radius R on a MoS$_2$ monolayer for the first few excited states; $(n=0,\ell=0)$, $(n=1,\ell=-1)$, $(n=1,\ell=0)$ and $(n=1,\ell=1)$.}\label{fig:Bc_r}
\end{figure}

\begin{figure*}[]
\centering
\subfigure[]{\includegraphics[width=0.49\linewidth]{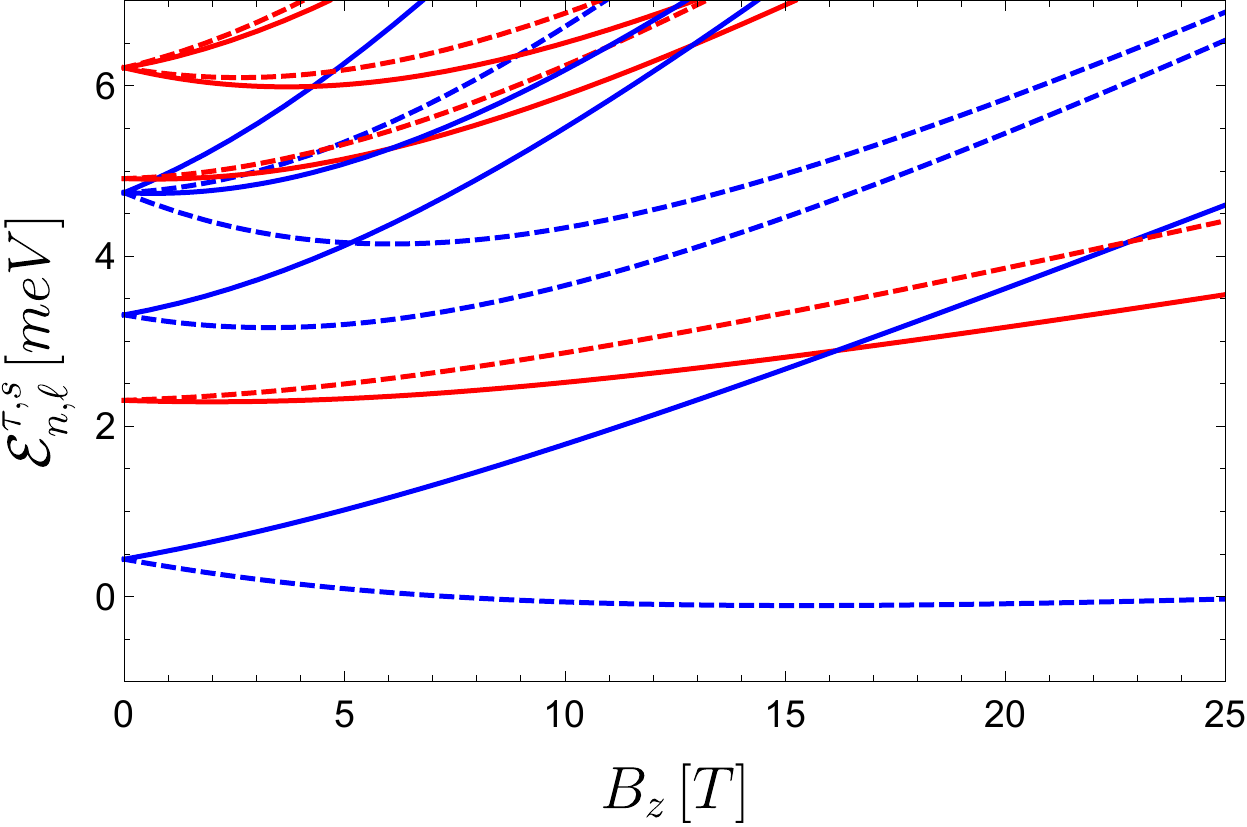}\label{fdplot}
}
\subfigure[]{\includegraphics[width=0.49\linewidth]{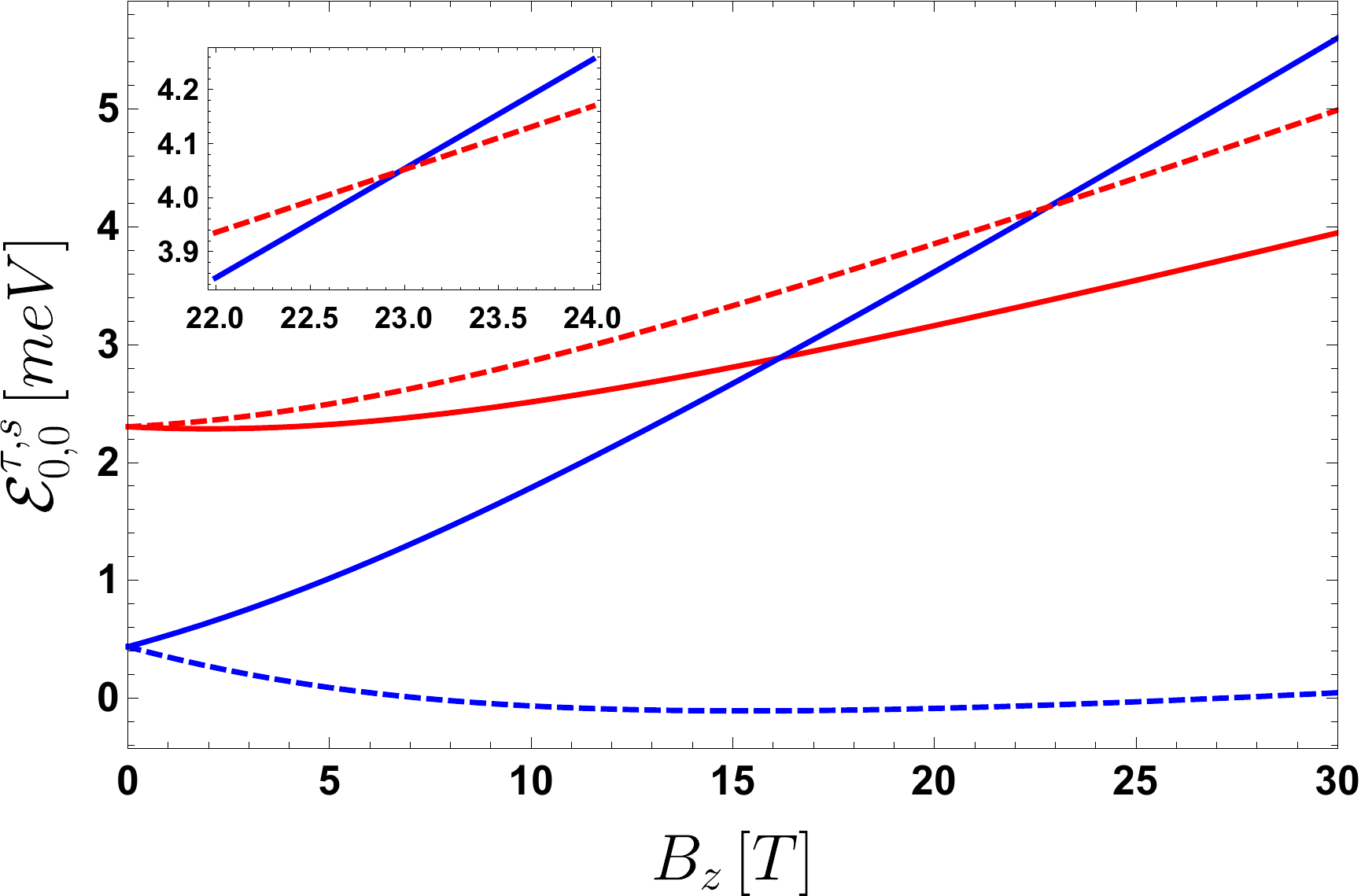}\label{fig:Groundstat}
}
\subfigure[]{\includegraphics[width=0.5\linewidth]{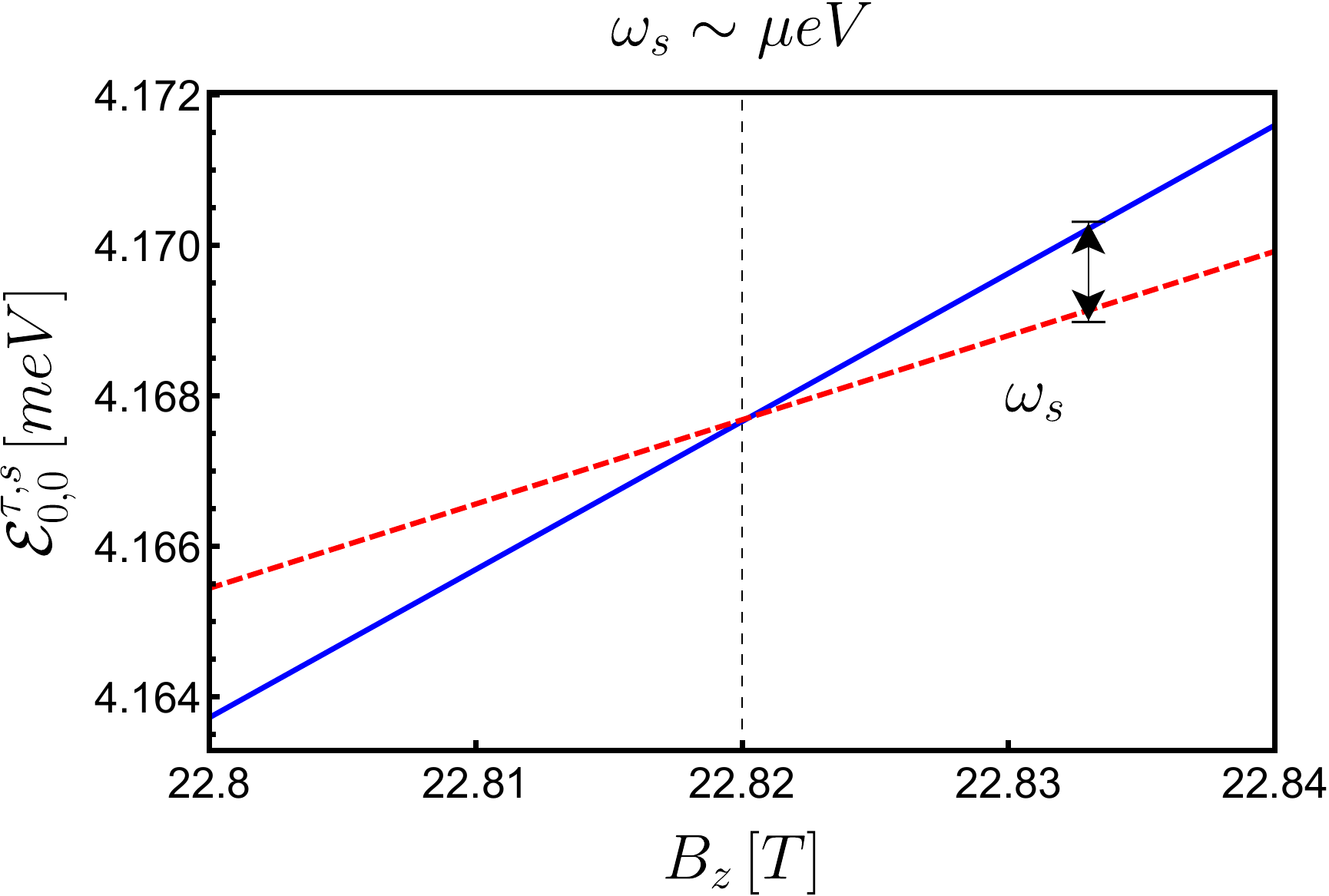} \includegraphics[width=0.5\linewidth]{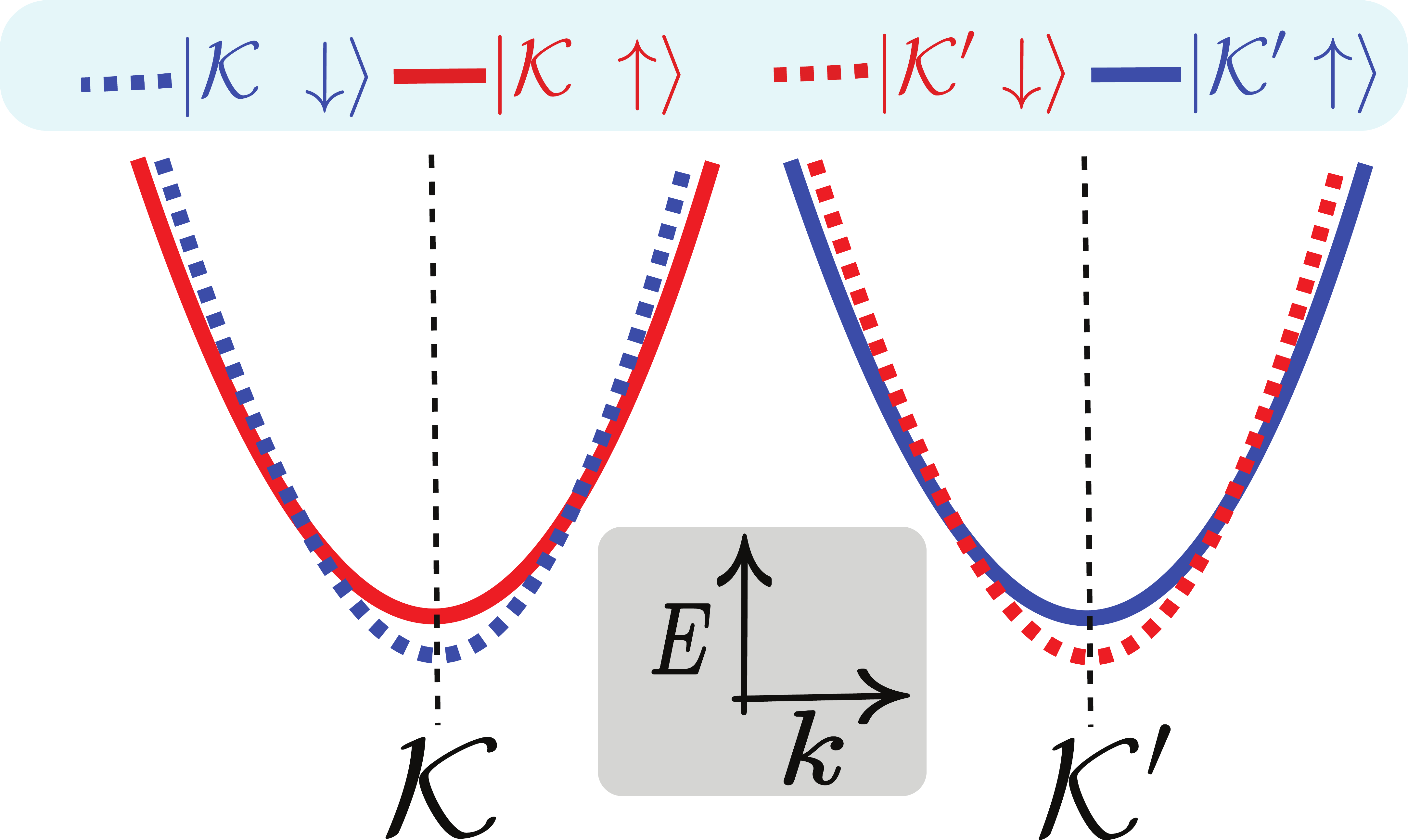}\label{fig:zoom_in} }
\caption{(a) Fock-Darwin Spectrum of a MoS$_2$ QD of radius $R = 26 \,nm$ as a function of the perpendicular magnetic field $B_z$. Blue solid line (dotted):  $\ket {\mathcal{K}^{\prime} \uparrow } (\ket { \mathcal{K} \downarrow})$ and continuous red line (dotted): $\ket { \mathcal{K} \uparrow} (\ket { \mathcal{K}^{\prime}\downarrow})$. States up to $|\ell|$ = 1 and $n = 1$ are shown. (b) Part of the spectrum shown in (a) represent the ground state energy spectra for $n = 0$, $\ell$ = 0 with a quantum dot radius $R = 26 nm$ of a MoS$_2$ monolayer experience a perpendicular magnetic field $B_z$. Inset: region about which the twofold degeneracy localized to valley $\mathcal{K}^{\prime}$ is observed in the spectrum. (c) Part of the spectrum shown in (b) with zoom in into  the twofold degeneracy of the two state $\ket { \mathcal{K}^{\prime}\downarrow}$ and $\ket { \mathcal{K}^{\prime}\uparrow}$.}
\label{image:E(B)}
\end{figure*}

% Conditions To be a TMD spin valley Qubit
To obtain a pure spin qubit in a single TMD QD, a thoughtful selection of parameters is necessary to acquire some robustness. Therefore, by selecting the appropriate TMD type, QD size and perpendicular magnetic field a regime where $ \mathcal{E}_{n,\ell}^{{\mathcal{K}(\mathcal{K}^{\prime}),\downarrow}} = \mathcal{E}_{n,\ell}^{{\mathcal{K}(\mathcal{K}^{\prime}),\uparrow}} $ may be achieved \cite{Brooks_2017}. MoS$_2$ is the semiconducting TMD monolayer with the smallest zero field spin splitting in the conduction band $\Delta_{cb}\simeq 2$ meV such that the condition $ \mathcal{E}_{n,\ell}^{{\mathcal{K}^{\prime},\downarrow}} = \mathcal{E}_{n,\ell}^{{\mathcal{K}^{\prime},\uparrow}} $ may be achieved for critical field. This critical magnetic field B$_z^{c}$ may be determined for a range of different QD radii to give the spin-degenerate regime $ \mathcal{E}_{n,\ell}^{{\mathcal{K}^{\prime},\downarrow}} = \mathcal{E}_{n,\ell}^{{\mathcal{K}^{\prime},\uparrow}}$, shown in Fig.~\eqref{fig:Bc_r}.  Remarkably, for QD radius $R \ge 26\, nm$ the value of $B_z^{c}$ stabilize to some equilibrium value $B_z^{c} = 22.82$ T for the ground state (n=0, $\ell$=0). Therefore, we assume $R = 26$ nm for the next step in this work. These spectra also show separate plateaus in the critical field strength at relatively high QD radii $R \ge 26 \,nm$ between the ground state (n=0, $\ell=0$) and the first excited states (n = 0, $\ell \ge 0$), differing by up to $\sim$ 5 T.\\
Fig.~\eqref{image:E(B)} depicts the numerically calculated energy spectrum of Fock-Darwin states for a QD with R = 26 nm in MoS$_2$. Fig.~\eqref{fig:Groundstat} shows the energy spectra of the ground state $\psi^{\tau,s}_{0,0}(r,\phi)$. In the absence of a magnetic field, we have two separate states due to the two different effective masses of electron in quantum dot. The magnetic field raises the degeneration of the levels into taking into account the degeneration of spin and valley such that the eigenbasis is described by the Kramers pairs  $\ket{\mathcal{K}^{\prime} \uparrow}, \ket{\mathcal{K} \downarrow}$ and $\ket{\mathcal{K} \uparrow}, \ket{\mathcal{K}^{\prime} \downarrow}.$ The Fig. \eqref{fig:zoom_in} shows the regime where $ \mathcal{E}_{n,\ell}^{{\mathcal{K}^{\prime},\downarrow}} = \mathcal{E}_{n,\ell}^{{\mathcal{K}^{\prime},\uparrow}} = 4.63\,$meV for critical magnetic
field strength $B_z^{c} = 22.82$ T which demonstrates spin degenerate crosses for a given radius in the $\mathcal{K}^{\prime}$ valley that we mentioned in previous requirements for the selection of a proper pure spin qubit.

With the intention of investigating the decoherence effect that occurs by hyperfine interaction with a single spin qubit, we consider a localized electron spin and in order to get the Zeeman splitting $\omega_s$ between the two state $\ket{0}= \ket{\mathcal{K}^{\prime},\uparrow}$ and 
$\ket{1}= \ket{\mathcal{K}^{\prime},\downarrow}$, we go slightly nearby  to this point of interaction and adding a small correction to the magnetic field,  $B_z^{c} + b_z$, where $b_z\sim 1.5\times10^{-2}$T. Therefore, $\omega_s=g_{sp}\mu_{B} b_z \simeq 1.7\, \mu$eV $\sim \text { hyperfine constant A }$  that's we will fit the problem for the next step of this work.\\
With a suitable operational regime selected, a TMD spin qubit has been theoretically demonstrated. Thus, the next step is to propose an approach to study the decoherence outcome that occurs by hyperfine interaction with a single spin qubit.
\section{Master Equation of an Electron Spin in MoS$_2$ Quantum Dot}\label{ME}
Electron spins can be manipulated via external controls by applying an external magnetic field and have been used as qubits, as shown in Sec.~\eqref{Qubit_candidat}. For this case, the electron wavefunction is localized inside a quantum dot with $R = 26$ nm. The effective Hamiltonian for this single electron spin interacts with a bath of a $\mathcal{N}$ spin-I$_0$ nuclei, through the contact hyperfine (HF) interaction, in a magnetic field b$_z$ along the z-axis is, ( setting $\hbar = 1$), 
\begin{equation} \label{ham_tot_gene}
\mathcal{H}_{\mathrm{tot}}=\omega_{s} S_{z}+ \omega_{k} I_z+ \mathbf{h}.\mathbf{S}
\end{equation}
where $\mathbf{S}=\left(S_{x}, S_{y}, S_{z}\right)$ is the electron spin operator. $\omega_s=g_{sp}\mu_{B} b_z$ $\left(\omega_{k} = g_{I_{k}} \mu_{N} b_z\right)$ is the electron (nuclear) Zeeman splitting in a magnetic field $b_z$, with effective $g$-factor $g_{sp}\left(g_{I}\right)$ for the electron (nuclei) and Bohr (nuclear) magneton $\mu_{B}\sim 2000 \,\mu_{N} \,\left(\mu_{N}\right)$, see Table~\eqref{table:mos2_nuc_gf}. For magnetic field $b_z\sim 1.5\times10^{-2}$T, $\omega_s=1.7\,\mu eV$.

\begin{figure}[h!]
\includegraphics[width=\linewidth ]{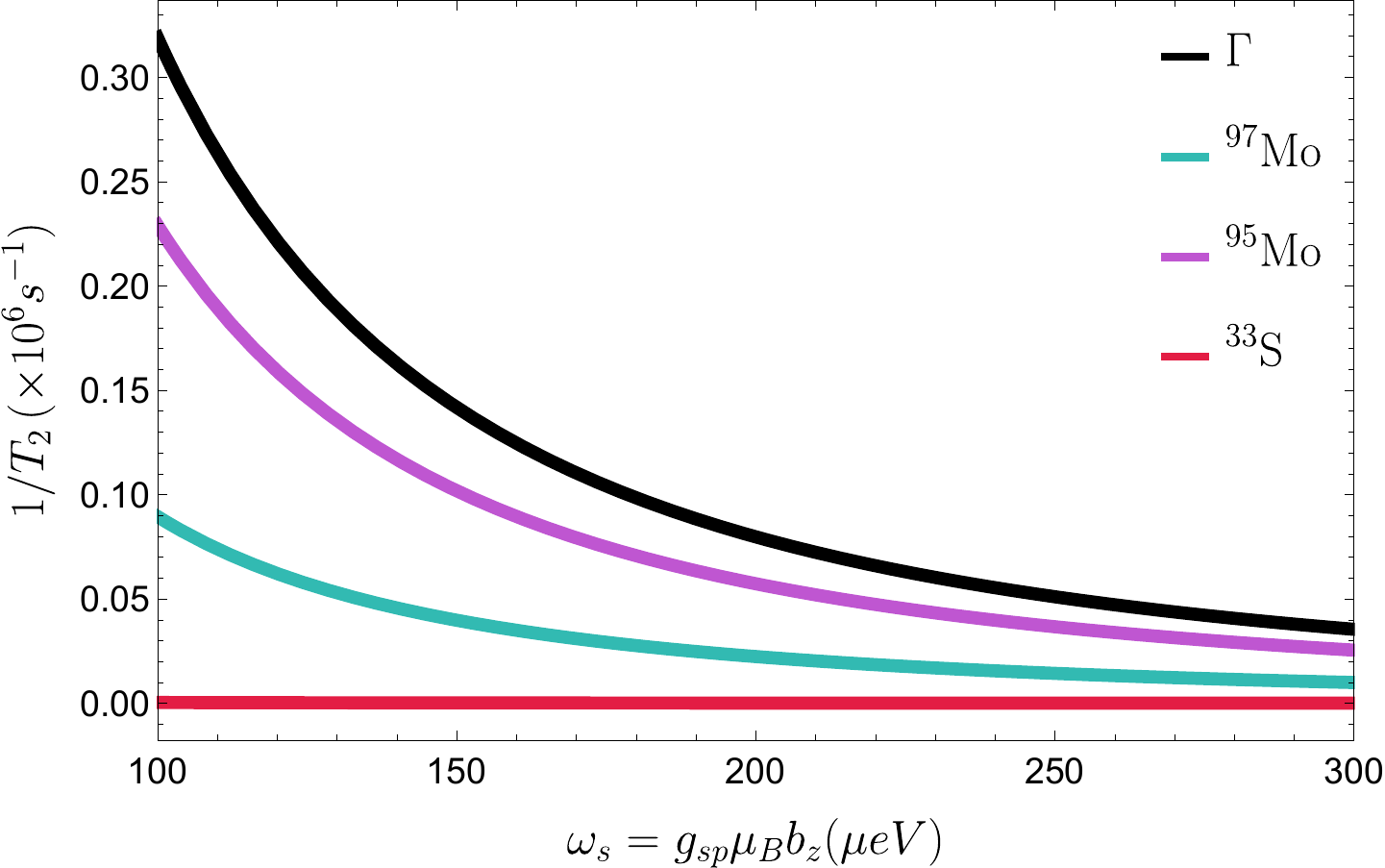} 
\caption{Decay rates for an MoS$_2$ quantum dot. Here, we have $\mathcal{N} = 10^4$ and used values of the abundance $\nu_{i}$ and the hyperfine constants $A^{i}$ for MoS$_2$ taken from the Table.~\eqref{tab:my_label}.}\label{decay_rate_t2}
\end{figure}

\begin{table}[]%The best place to locate the table environment is directly after its first reference in text
\caption{\label{table:mos2_nuc_gf} Nuclear g-factor for different isotopes, that allowing non zero nuclear spin, and nuclear Zeeman splitting in a magnetic field $b_z\sim 1.5\times10^{-2}$T for MoS$_2$. Note that $\mid\omega_s\mid = (10^3 \sim 10^4)\mid\omega_k\mid$, as for the nuclear spin frequencies $\omega_k$, their magnitudes are three-order smaller than $\omega_s$.
}
\begin{ruledtabular}
\begin{tabular}{cccc}
Isotopes&$^{95}$Mo&$^{97}$Mo&$^{33}$S\\ \hline
g$_I$&-0.3657&-0.3734&0.4292\\ 
$\mid\omega_k\mid=\mid g_{I}\mu_{N} b_z\mid$  [$10^{-3}\times\mu eV$]&0.17&0.18&0.2\\
\end{tabular}
\end{ruledtabular}
\end{table}
In Eq.~\eqref{ham_tot_gene} we have neglected the anisotropic hyperfine interaction, electron-electron interaction, dipole-dipole interaction between nuclear spins, and nuclear quadrupolar splitting, which may be present for nuclear spin I$_0 > 1/2$~\cite{Coish}. The total action by an environment of the $\mathcal{N}$ nuclear spins can be interpreted as a nuclear magnetic field, is the so-called Overhauser field \cite{Sarmaarticle}
\begin{equation}
    \mathbf{h}=\left(h_{x}, h_{y}, h_{z}\right)=\sum_{k=0}^{\mathcal{N}-1} A_{k} \mathbf{I}_{k}
\end{equation}
where $\mathbf{I}_{k}$ $=\left(I_{k}^{x}, I_{k}^{y}, I_{k}^{z}\right)$ is the nuclear spin operator at lattice site $k$ at position $\textbf{r}_{k}$. $I_{z}=\Sigma_{k} I_{k}^{z}$ is the total $z$ component of nuclear spin and  $A_{k}$ is the associated hyperfine coupling constant. We have also introduced raising and lowering operators $S_{\pm}=S_{x} \pm i S_{y}$,  $I_{k}^{\pm}=I_{k}^{x} \pm i I_{k}^{y}$ and the nuclear magnetic field operators $h^{\pm}=h_{x} \pm i h_{y}$. We can rewrite the Eq. \eqref{ham_tot_gene}
\begin{align}\label{modified}
\mathcal{H}_{\mathrm{tot}}=\omega_{s} S_{z}+&\sum_{k} \omega_{k} I_{k}^{z}+\sum_{k} \frac{A_{k}}{2}\left(S_{+} I_{k}^{-}+S_{-} I_{k}^{+}\right)\\&+\sum_{k} A_{k} S_{z} I_{k}^{z}\notag
\end{align}
The third and the fourth terms in Eq. \eqref{modified} is the hyperfine contact interaction between the spin electron and the nuclei in the quantum dot, which describe the flip-flop interaction and (longitudinal) Overhauser’s field, giving rise to inhomogeneous broadening and dephasing, indeed the last terme, $\sum_{k}A_{k } S_zI_{z}^{k}$, produces an effective magnetic field for the electron $B_{\text {eff }}=b_{z}-\mathcal{B}_N$, where~\cite{urbaszek2013nuclear}
 \begin{equation}
 \mathcal{B}_N = \sum_{k} A_{k} I_{z}^{k}/ (g_{sp} \mu_{B})
 \end{equation}
which results in the well known Overhauser shift. However, when $g_{sp} \mu_{B} B_{\text {eff }} \ll (\sum_{k} A_{k}/2\left(S_{+} I_{k}^{-}+S_{-} I_{k}^{+}\right))$, spin exchange becomes the dominate effect \cite{Taylor_2003}. The strength $A_k$ is determined by the electron density at the site of nuclei~\cite{schliemann2003electron}
 \begin{equation}
    A_{k}=A^{i_k} v_{0}\left|\psi_{0, 0}^{\mathcal{K^{'}}, s}\left(\mathbf{r}_{k}\right)\right|^{2}
\end{equation}
corresponds to the one-electron hyperfine interaction with the nuclear spin at site k with position ${r}_{k}$. Here, $v_{0}= \sqrt{3}\,a^{2}/4$ is the (two-dimensional) volume of a crystal unit cell containing one nucleus, a is the lattice parameter. Indeed, in QD with radius R includes $\mathcal{N}_{tot}=\pi R^{2}/v_{0}$ nuclei in total. For MoS$_{2}$ QD of size $R = 26 \,nm$ with lattice constant given by a = 3.19 $\AA$, there are $\mathcal{N}_{tot}\sim 10^{4}$ nuclei within the dot. However, only the isotopes that allowing non zero nuclear spin that's will part of these nuclei, $^{95}$Mo, $^{97}$Mo and $^{33}$S. Furthermore, the concentration of $^{33}$S is negligible compared to that of Mo isotopes, and the decoherence of the electron spin mainly originates from the presence of $^{95}$Mo and $^{97}$Mo nuclear spins~\cite{ye2019spin}. This leads to the number of  nuclear spins within the QD, $\mathcal{N}=(\sum_{i}\nu_i)\mathcal{N}_{tot}$, where $\nu_i$ is the natural abundance for different nuclear isotopic species $i$. $\psi_{0, 0}^{\mathcal{K^{'}}, s}(\mathbf{r_k})$ is the envelope wave function of the localized electron, and $A^{i_k}$ is the total hyperfine coupling constant to a nuclear spin of species $i_k$ at site k, where is given by \cite{Sarmaarticle},
\begin{equation}
A^{i_{k}}=-\frac{\mu_{0}}{4 \pi} \cdot \frac{8 \pi}{3} \gamma_{S} \gamma_{i_{k}}\left|u_{i_{k}}\right|^{2}
\end{equation}
where, $\mu_{0}$ is the vacuum permeability. $u_{i_{k}}$ is the amplitude of the periodic part of the Bloch function at the position of the nucleus of $i_{k}$ species. $\gamma_S$  is the gyromagnetic ratio of free electrons, and its value is always negative. However,  $\gamma_{i_{k}}$ which is the nuclear gyromagnetic ratio can take either sign. This fact causes the hyperfine coupling constant $A^{i_{k}}$ to be either positive or negative. For convenience, in a material containing several different nuclear isotopic species $i_k$ we define an average hyperfine coupling constant. Here, we take the Root mean square(RMS) average~\cite{Coish,Coish2009}
\begin{equation}\label{A_hf}
A=\sqrt{\sum_{i} \nu_{i}\left(A^{i}\right)^{2}}
\end{equation}
For MoS$_2$, the hyperfine constant estimated for $^{95}$Mo and $^{97}$Mo is $A^{^{95}\text{Mo}}$=$A^{^{97}\text{Mo}}$ = - 0.57 $\mu$eV \cite{avdeev2019hyperfine}, by using these coupling constants with the abundances listed in Table~\eqref{tab:my_label} gives an RMS  coupling strength A = 0.29 $\mu$eV. To be more efficient, in order to show the isotopes that have the main contrubition to the decoherence due to the hyperfine interaction in MoS$_2$, we plot the decoherence rate  $1/T_2 = \Gamma = \sum_i \Gamma_i$, See Fig.~\ref{decay_rate_t2}, 
may be defined as~\cite{Coish}
\begin{equation}\label{eq:decayT2}
\Gamma_{i}=\frac{1}{T_{2}^{i}}=\nu_{i}^{2} \frac{\pi}{3}\left(\frac{I_{i}\left(I_{i}+1\right) A^{i}}{3 \,\omega_s}\right)^{2} \frac{A^{i}}{\mathcal{N}}
\end{equation}
where $\Gamma_{i}$ is the contribution from flip-flops between nuclei of the common species i. The quadratic dependence on isotopic abundance $\nu_{i}$, shown in Eq.~\eqref{eq:decayT2}, is particularly an important factor in the decoherence rate. Due to this dependence, electron spins in MoS$_2$, where Mo has two naturally occurring isotopic species, whereas S has only one, will show a decay mostly due to flip-flops between Mo spins, $^{95}$Mo notably. Significantly, we note that the relatively large flip-flop rates for Mo isotopes, as a result of a large nuclear spin $5/2$ and isotopic natural abundance, respectively.
Now, we consider a localized electron in its orbital ground state. This can be seen, for instance, the wavefunction for the ground state of a single electron isolated in the conduction band (CB) about the valley $\mathcal{K}$ and $\mathcal{K^{'}}$ under a magnetic field for a quantum dot (QD), with parabolic confinement, in monolayer MoS$_2$, which is given by,
\begin{equation}\label{wf}
\psi_{0, 0}^{\mathcal{K^{'}}, s}(r_k) = \frac{1}{\sqrt[4]{\pi}\ell_0} exp(-\frac{1}{2}(\frac{r_k}{\ell_0})^{2})\chi_s(\sigma_s)
\end{equation}
where, $\ell_0$ is the effective length scale, equal to the magnetic length $\ell_B = (\hbar /(e B_{z}))^2$ in the absence of the confining potential ($\omega_0 \rightarrow 0$), giving by,
\begin{equation}
 \ell_0 = \sqrt{\frac{\hbar}{2 m_{eff}^{\mathcal{K^{'}}, s} \Omega_{\mathcal{K^{'}}, s}}}
\end{equation}

Due to our selection in Sec.~\eqref{Qubit_candidat} for the desired spin qubit, $\tau$ would take the value -1 (valley $\mathcal{K}^{\prime}$), therefore the ground state wavefunction that will be considerd for this work is $\psi^{\mathcal{K}^{\prime}, s}_{0,0}(r_k)$.

Adding and subtracting $\sum_k A_k I_k^{z}/2$ to the total Hamiltonian 
\begin{widetext}
\begin{equation}\label{Htot_}
\begin{split}
\mathcal{H}_{\mathrm{tot}}&=\omega_{s} S_{z}+\sum_{k} \omega_{k} I_{k}^{z}+\sum_{k} \frac{A_{k}}{2}\left(S_{+} I_{k}^{-}+S_{-} I_{k}^{+}\right)+\sum_{k} A_{k} S_{z} I_{k}^{z} + \sum_k \frac{A_k}{2} I_k^{z}
 -\sum_k \frac{A_k}{2} I_k^{z}\\
&=\underbrace{\omega_{s} S_{z}+\sum_{k} ( \omega_{k} - \frac{A_k}{2})I_{k}^{z}}_{\mathcal{H}_0}  +\underbrace{\sum_{k} \frac{A_{k}}{2}\left(S_{+} I_{k}^{-}+S_{-} I_{k}^{+}\right)+\sum_{k} A_{k} (S_{z}+\frac{1}{2}) I_{k}^{z}}_{\mathcal{H}_1}\\
&=\mathcal{H}_0 + \mathcal{H}_1
\end{split}
\end{equation}
\end{widetext}
This Hamiltonian \eqref{Htot_} can be split in two main part, an unperturbed part (longitudinal) noted $\mathcal{H}_0$ which consisting of all Zeeman terms and a perturbative part (transverse) $\mathcal{H}_1$ containing the virtual flip-flop processes of the hyperfine interaction (HI).\\
In the interaction picture with respect to the Hamiltonian $\mathcal{H}_0$
\begin{align}\label{TL}
\mathcal{H}_{tot}^{I}&= e^{i \mathcal{H}_0 t } \mathcal{H}_1 e^{-i \mathcal{H}_0 t }\\
&= \underbrace{ S_{+}(t) h^{-}(t)+S_{-}(t) h^{+}(t)}_{\text{Transversal Hyperfine Term}}+\underbrace{|1\rangle\langle 1| h^{z}}_{\text{Longitudinal}}\notag
\end{align}
where
\begin{align}
  &h^{\pm}(t) = \sum_{k} \frac{A_{k}}{2} I_{k}^{\pm} e^{\pm i \left(\omega_{k}-A_{k} / 2\right) t},\\&S_{\pm}(t)=S_{\pm} e^{\pm i \omega_{s} t}, \notag\\& h^{z} = \sum_{k} A_{k} I_{k}^{z}  \notag
\end{align}
Note that the last term in Eq.~\eqref{TL}  is equivalent to $\sum_{k} A_{k} (S_{z}+\frac{1}{2}\mathds{1}_2) I_{k}^{z}$, where $S_{z} = (|1\rangle\langle 1|-| 0\rangle\langle 0|) / 2$, with $|1\rangle=\left|\mathcal{K}^{\prime} \downarrow\right\rangle (|0\rangle= \left|\mathcal{K}^{\prime} \uparrow\right\rangle)$ is the down (upper) state of the central spin. Indeed, $S_{z} + \frac{1}{2} \mathds{1}_2 = |1\rangle\langle 1|$. $\mathds{1}_2$ is the $2 \times 2$ identity matrix.
The transversal hyperfine term results in the off-resonant transitions between the system and environmental spins, while the longitudinal hyperfine term provides additional contributions to the energy splitting.

The time evolution of the combined system, consisting of the electron spin and $\mathcal{N}$ nuclear spins, which is given by the action of the total Hamiltonian $\mathcal{H}_{tot}^{I}$ in Eq.~\eqref{TL}, is described as the following. To begin with, when $t < 0$ we assume that the electron spin and the nuclear system are decoupled, and both of them prepared 
 independently in the states described by the density operators $\rho_{S}(0)$ and $\rho_{E}(0)$, respectively.
At $t$ $=0,$ the electron and nuclear spin system are brought into contact over a switching time scale $ \tau_{s w} \ll 2 \pi \hbar/\left|\omega_s-\omega_k+A\right|$~\cite{coish2004hyperfine}, which is sufficiently small, where $\left|\omega_s-\omega_k+A\right|$ is the largest energy scale in this problem. The state of the entire system, described by the total density operator $\rho(t)$, where it's given at $t=0$, 
\begin{equation}\label{density_matrix_intial}
\rho(0)=\rho_{S}(0) \otimes \rho_{E}(0)
\end{equation}with $\rho_{S} = \operatorname{Tr}_{E}(\rho)$ and $\rho_{E} = \operatorname{Tr}_{S}(\rho)$, this is called the reduced density matrix of the subsystem S and E, respectively. The evolution of the density operator $\rho(t)$ for $t \ge 0$ is governed by the Hamiltonian $\mathcal{H}^{I}_{tot}(t)$ for an electron spin coupled to an environment of nuclear spins.

This model that we consider in this work is similar to previous research studies of an electron spin confined to a Gallium arsenide(GaAs) QD \cite{coish2004hyperfine} and Graphene QD \cite{fuchs2012spin}, but there are  MoS$_2$ specific properties, that lead to new physics. Given that the natural abundance $\nu_i$ of spin-carrying isotopes is small for molybdenum Mo and sulfur S, hence only $\mathcal{N}$  of all atoms $N_{tot}$ within the MoS$_2$ QD carry spin. However,  in semiconducting materials such as GaAs, all isotopes possess a spin. To highlight these differences, we do a comparison of the most important characteristics of MoS$_2$, graphene, and GaAs which is given in Table \eqref{tab:my_label}. Indeed, the hyperfine interaction coupling constant A$_{\text{MoS$_2$}}$, is about approximately the same magnitude as  A$_{\text{Graphene}}$  and it's about two orders of magnitude smaller than the constant A$_{\text{GaAs}}$ in GaAs which even further reduce the nuclear magnetic field by the same amount. Moreover, The relatively small hyperfine energy, $\mathcal{E}_{HF}=\hbar/\tau_{HF}$ (in case of MoS$_2$ is on average of order $\mathcal{E}_{HF}\thickapprox 10^{-12}$eV), in MoS$_2$ increases the timescale $\tau_{HF}$, which is depending on both the hyperfine strength A and the isotopic abundance $\nu_{i}$, of this interaction significantly as compared to GaAs and graphene. The switching time scale $\tau_{sw}$ for several materials are listed in Table \eqref{tab:my_label} as well.
\begin{table*}[]
\caption{\label{tab:my_label}Comparison of the most important parameters of GaAs, Graphene($^{13}\mathbf{C}$) and MoS$_2$. The main isotopes that allowing non zero nuclear spin. The total number of nuclei $\mathcal{N}_{\text{tot }}$ estimated for a QD of typical size $R = 26 \,nm$. The hyperfine strength A and the time scale $\tau_{HF} \sim 2\hbar\mathcal{N}/\nu A$~\cite{avdeev2019hyperfine, fuchs2012spin} for the decay of the electron spin due to the contact hyperfine interaction.}
\begin{ruledtabular}
\renewcommand{\arraystretch}{1.2}
\begin{tabular}{cccccccccccc}
    & Units & \multicolumn{3}{c}{GaAs}& $^{13}\mathbf{C}$&\multicolumn{3}{c}{MoS$_2$}\\
    \hline
    Main Isotopes & {[1]} & ${}^{71}$Ga& ${}^{69}$Ga& ${}^{75}$As&$ {}^{13}$C& ${}^{95}$Mo &$^{97}$Mo &$^{33}$S\\
    $\mathcal{N}_{\text {tot }}$ & {[1]} &  &$15\times 10^{3}$&& $8\times10^{4}$&&$5\times 10^{4}$&\\
    Abundance $\nu_{i}$ & {[1]} & 39.89\%& 60.10\%& 100\%& $1.07\% $& 15.92\%& 9.55\%& 0.76\%\\
    $\mathcal{N}$ & {[1]} & &$15\times10^{3}$& & $8\times10^{2}$&& $13\times10^{3}$& \\
    $A^{i}$ & $[\mu e V]$ &96\cite{Coish2009}&74\cite{Coish2009}& 86\cite{Coish2009}& 0.6\cite{fuchs2012spin}&-0.57\cite{avdeev2019hyperfine}&-0.57\cite{avdeev2019hyperfine}& 0.75\cite{avdeev2019hyperfine}\\
    I$_0$ & {[1]} &3/2& 3/2 &3/2& 1/2&5/2&5/2&3/2 \\
    $\tau_{H F\propto \hbar/A}$ & $[\mu s]$ & &0.2& & 178&&219& \\
    $\tau_{SW}\sim 2\pi\hbar/A$ & [ns]& &$ 5.10^{-2} $& & $7$& &$14$&\\
\end{tabular}
\end{ruledtabular}
\end{table*}

\begin{figure*}[]
\subfigure[]{\includegraphics[width=0.49\linewidth]{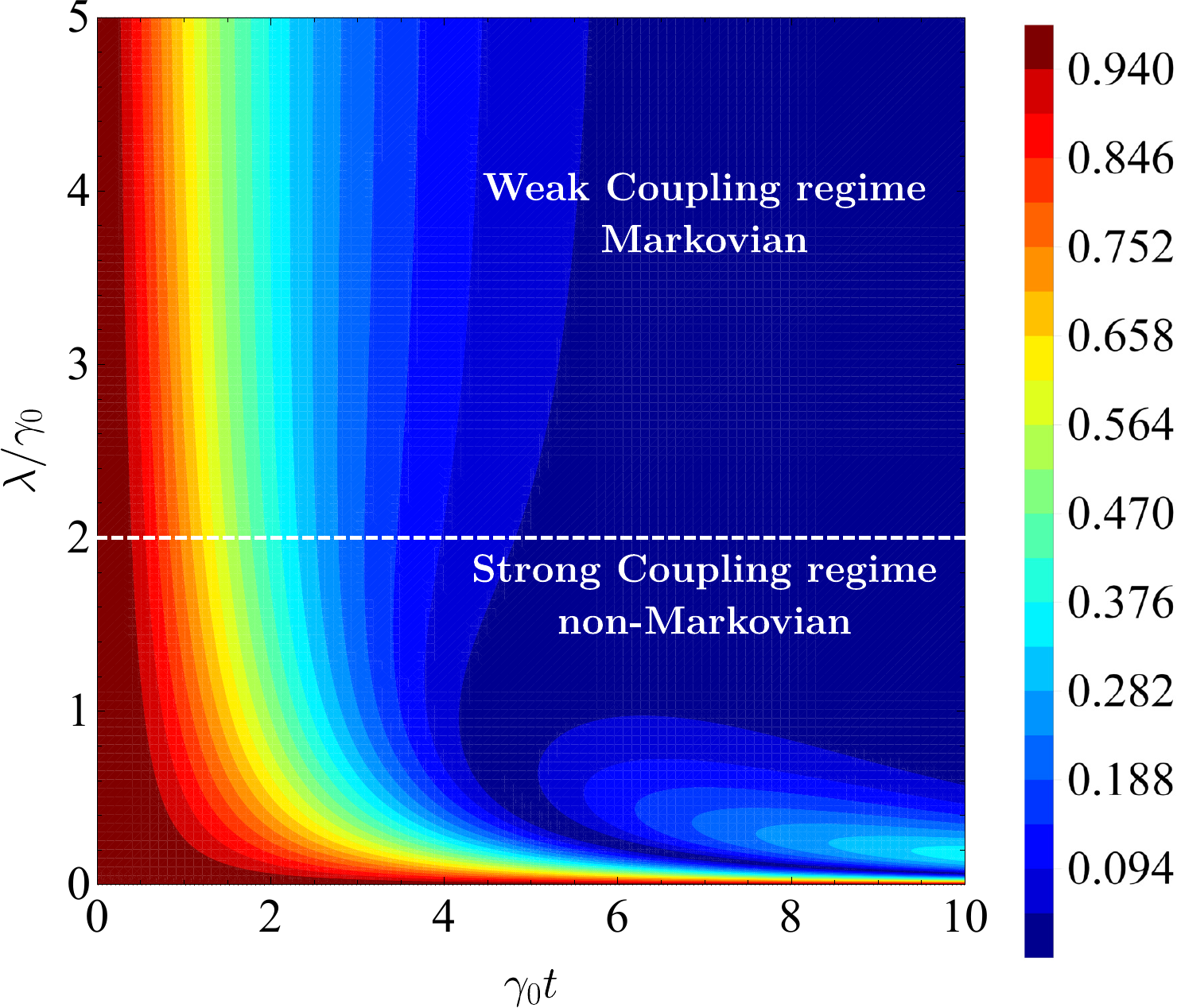} \label{Fidlity_general}}
\subfigure[]{\includegraphics[width=0.49\linewidth,height=7cm]{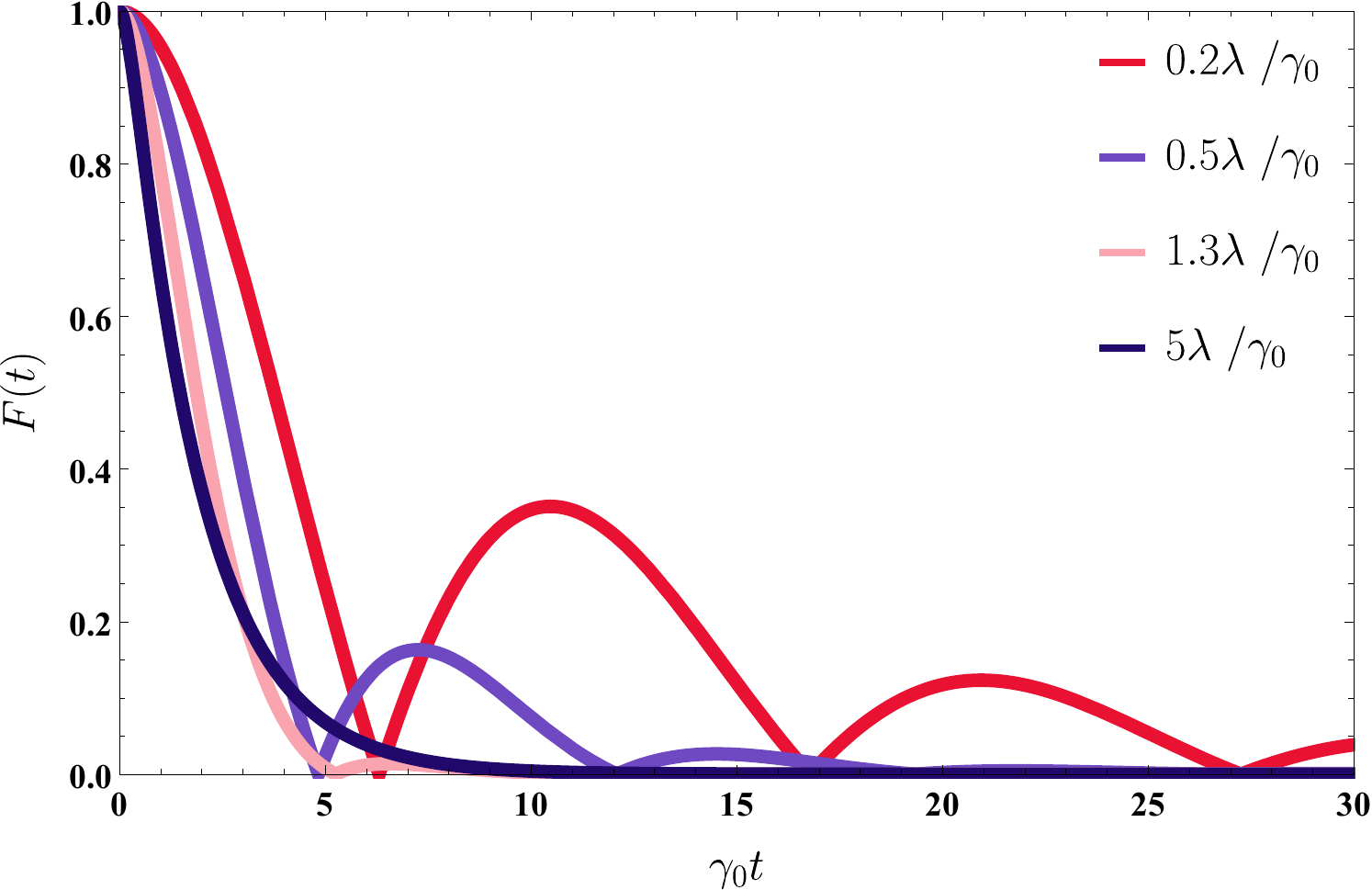}\label{Fidlity_gamma_0_t} }
\caption{Time evolution of the fidelity $\mathcal{F}(t)$ in Eq. \eqref{fid} using G(t) in Eq. \eqref{propagator_g}, (a) as a function of dimensionless time $\gamma_0\,t$ and environmental memory parameter $\lambda/\gamma_0$, (b) as a function of dimensionless time $\gamma_0\,t$ for a different values of the environmental memory parameter $\lambda/\gamma_0$.}
\end{figure*}

Next, We drive into the method that used along side with this work.
We First suppose at $t=0$, that the total system (electron and nuclear) describe by
\begin{equation}\label{stat_tot}
    \psi(0)= \psi_S(0)\otimes\psi_E(0)
\end{equation}
\begin{figure}[h!]
    \centering
    \includegraphics[width=\linewidth]{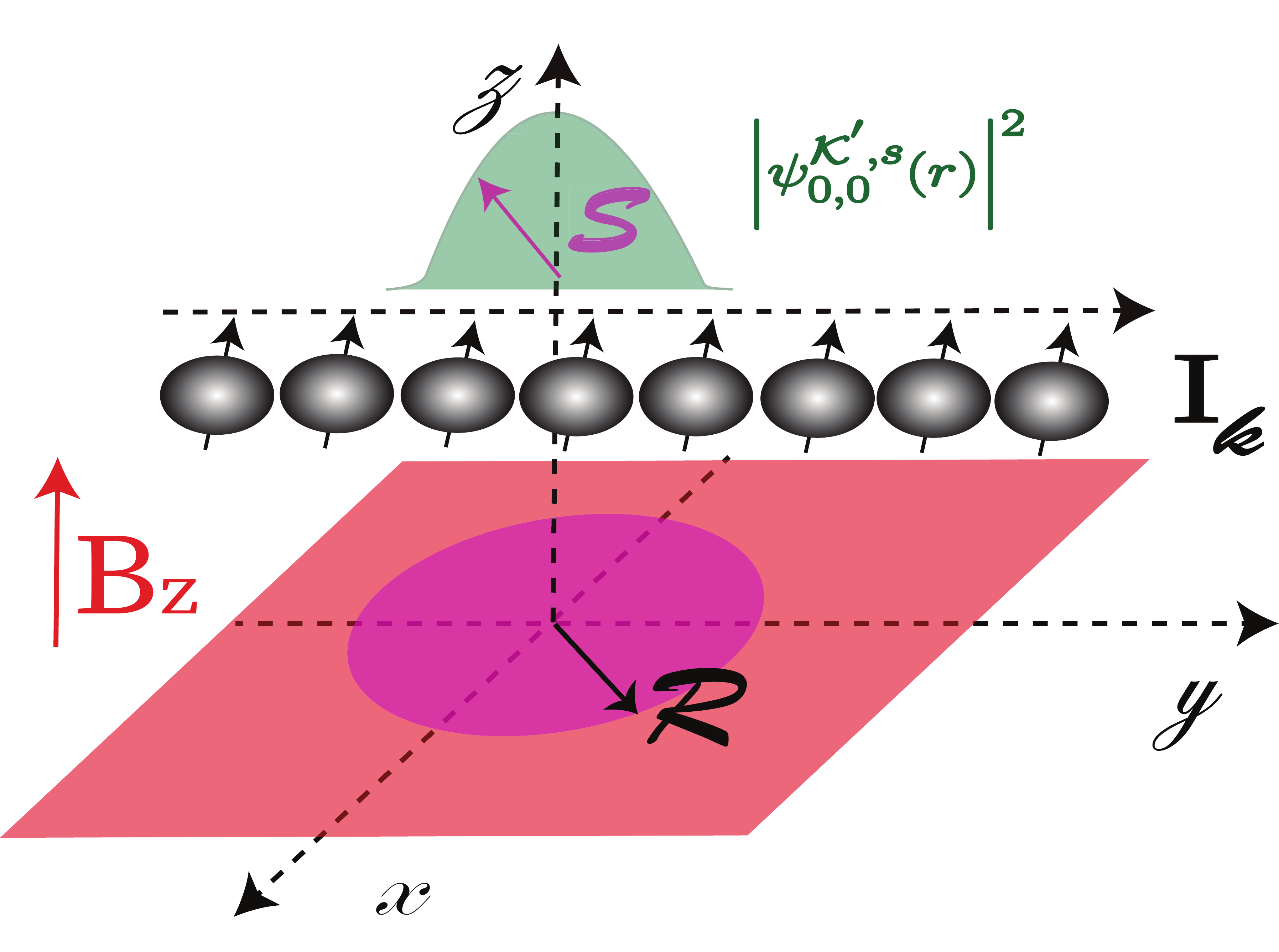} 
    \caption{  Schematic illustration of a two-dimensional MoS$_2$ quantum dot (QD). An electron spin S is localized in orbital ground state of the QD, where it is immersed with a full polarized bath of nuclear spins $I_{k}$. Due to the confinement, the spatial distribution of the electron in the ground state of the QD is described by a Gaussian envelope function $\psi_{0, 0}^{\mathcal{K^{'}}, s}(r)$ given in Eq. \eqref{wf}, which in turn leads to a non-uniform hyperfine interaction between the electron spin and the nuclear spins.  An external magnetic field is applied perpendicular ($B_{z}$) to the plane of the dot (x,y). }
    \label{intial_stat_sys}
\end{figure}
where $\ket{\psi_S(0)}$, is the initial state for the central spin and $\ket{\psi_E(0)}$ is the initial state for the nuclear spin bath where we start with a perfectly polarized nuclear ensemble, as shown in Fig. \eqref{intial_stat_sys}. The Eq. \eqref{stat_tot} can be written in the subspace spanned by the bases $\lbrace |0\rangle\otimes|0\rangle_E, |1\rangle\otimes|0\rangle_E,|0\rangle \otimes \left( I^{+}_{k}|0\rangle_E\right)\rbrace$ as following:
\begin{align}
    \mid\psi(0)\rangle=&c_0 |0\rangle\otimes|0\rangle_E+ c_1(0) |1\rangle\otimes|0\rangle_E\\&+ \sum_k c_{k}(0) |0\rangle \otimes \left( I^{+}_{k}|0\rangle_E\right) \notag
\end{align}
where, $|0\rangle_E =\otimes_{k}^{{\mathcal{N}}}|0\rangle= |\underbrace{ 0, \cdots,0}_{{\mathcal{N}}}\rangle  $ denote the vacuum state of the bath. Noting that, $c_{k}(0) = 0\,\forall\,k$. This means
that the environment is in the vacuum state initially. The time evolution of the total system, $t>0$, can be written by the following expression
\begin{align}\label{psi_t}
    \mid\psi(t)\rangle=&c_0 |0\rangle\otimes|0\rangle_E+ c_1(t) |1\rangle\otimes|0\rangle_E\\&+ \sum_k c_{k}(t) |0\rangle \otimes \left( I^{+}_{k}|0\rangle_E\right) \notag
\end{align}
where we have used the normalization condition $\mid c_0\mid^2+\mid c_1(t)\mid^2+\sum_k\mid c_k(t)\mid^2=1$. Let us introduce the states 
\begin{equation}
\ket{\psi_{0}}=|0\rangle \otimes|0\rangle_{E},\:
\ket{\psi_{1}}=|1\rangle \otimes|0\rangle_{E},\:
\ket{\psi_{k}}=|0\rangle \otimes|k\rangle_{E}
\end{equation}
where $|k\rangle_{E}= I^{+}_{k}|0\rangle_E = \ket{0_1, \cdots, 0_{k-1},1_k,0_{k+1,\cdots}}$ denotes the state with only one nuclear spin in site k. We can now 
express the time evolved state \eqref{psi_t} as:
\begin{equation}
\mid\psi(t)\rangle=c_0 \ket{\psi_{0}}+ c_1(t) \ket{\psi_{1}}+ \sum_k c_{k}(t) \ket{\psi_{k} }
\end{equation}
The amplitude $c_{0}$ is constant since $\mathcal{H}^{I}_{tot}(t)\ket{ \psi_{0}}=0$, while the amplitudes $c_{1}(t)$ and $c_{k}(t)$ are time dependent. The time development of these amplitudes is governed by a system of differential equations which is obtained by the Hamiltonian in Eq. \eqref{TL} and the Schrödinger equation, 
\begin{equation}
\begin{split}
i \partial_{t}|\psi(t)\rangle & = \dot{c}_{1}(t)\left|\psi_{1}\right\rangle+\sum_{k} \dot{c}_{k}(t)\left|\psi_{k}\right\rangle\\
&= \mathcal{H}_{tot}^{I}|\psi(t)\rangle = S_{+}(t) h^{-}(t)+S_{-}(t) h^{+}(t)+|1\rangle\langle 1| h^{z}\\
&\times\left(c_{0}\left|\psi_{0}\right\rangle+c_{1}(t)\left|\psi_{1}\right\rangle+\sum_{k} c_{k}(t)\left|\psi_{k}\right\rangle\right) \\
& = \sum_{k} \frac{A_{k}}{2} e^{-i \left(\omega_{s}-\omega_{k}+A_{k} / 2\right) t} c_1(t) \ket{\psi_k}\\
&+ \left( h c_1(t)+ \sum_{k} c_k(t)\frac{A_{k}}{2} e^{-i \left(\omega_{k}-\omega_{s}-A_{k} / 2\right) t} \right) \ket{\psi_1}
\end{split}
\end{equation}
where $h=\left\langle h^{z}(t)\right\rangle=\left\langle\psi(t)\left|h_{z}\right| \psi(t)\right\rangle$. Multiplying by $\bra{\psi_1}$ and $\bra{\psi_k}$ gives us two coupled differential equations for the amplitudes $c_{k}(t)$ and $c_{1}(t)$:
 \begin{equation}\label{amplitude}
\begin{array}{c}
\frac{d}{d t} c_{1}(t)=i h c_{1}(t)-i \sum_{k} \frac{A_{k}}{2} e^{i\left(\omega_{s}-\omega_{k}+\frac{A_{k}}{2}\right) t} c_{k}(t) \\
\frac{d}{d t} c_{k}(t)=-i \frac{A_{k}}{2} e^{-i\left(\omega_{s}-\omega_{k}+\frac{A_{k}}{2}\right) t} c_{1}(t)
\end{array}
\end{equation}
By integrating Eq.~\eqref{amplitude}, the coefficient $c_{k}(t)$ can be formally written as
\begin{equation}
c_{k}(t)=-i \frac{A_{k}}{2} \int_{0}^{t} ds\, e^{-i\left(\omega_{s}-\omega_{k}+\frac{A_{k}}{2}\right) s} c_{1}(s)
\end{equation}
Substituting it into Eq.\eqref{amplitude}, we obtain an exact time-convolution dynamical equation for the central spin
\begin{equation}
\frac{d}{d t} c_{1}(t)=i h c_{1}(t)- \int_{0}^{t} d s \left\langle h^{-}(t) h^{+}(s)\right\rangle_{E} e^{i \omega_{s}(t-s)} c_{1}(s)
\end{equation}
where, $\left\langle h^{-}(t) h^{+}(s)\right\rangle_{E} e^{i \omega_{s}(t-s)} =\sum_{k}\left(A_{k}/2\right)^{2} e^{i\left(\omega_{s}-\omega_{k}+A_{k}/2\right)(t-s)}$ is the associated memory kernel function given by a two-point correlation function of the bath.  $\langle\cdots\rangle_{E}$, is the expectation values over the state of the environment $\rho_E$, where $\rho_E(t)=\left(\ket{0}\bra{0}\right)_E$ is the vacuum state of the bath. The correlation function in the continuum limit assumes the form 
$\int d\omega \mathcal{J}(\omega)\, e^{i \left(\omega_{s}+A_{k}/2 - \omega\right)(t-s) }$, with $\mathcal{J}(\omega)=\sum_k \left(A_{k}/2\right)^{2} \delta(\omega-\omega_k) $ is the spectral density of the bath given by sum of  (coupling strength)$^{2}$ × (density of modes), which is therefore simply the Fourier transform of the correlation function $\left\langle h^{-}(t) h^{+}(s)\right\rangle_{E} e^{i \omega_{s}(t-s)}$. The typical representative environments are described by the Lorentzian-type spectral functions, 
$
\mathcal{J}(\omega)=\gamma_{0} \lambda^{2}/\left(2 \pi\left(\omega_{s}-\omega\right)^{2}+\lambda^{2}\right)
$.
 Here, the parameter $\lambda$ defines the width of the Lorentzian spectral density, indeed its the measure of the memory capacity or non-Markovianity of the environment \cite{jing2018decoherence} and is connected to the bath correlation time $\tau_{E}=1/\lambda$. On the other hand, $\gamma_{0}$ measures the strength of coupling between the qubit and its environment and
hence the system characteristic time $\tau_{S}=1/\gamma_{0}$ denotes the relaxation time. By taking a Lorentzian spectral density in resonance with the transition frequency of the qubit we find an exponential two-point correlation function, expressed as
\begin{equation}\label{General_f}
\left\langle h^{-}(t) h^{+}(s)\right\rangle_{E} e^{i \omega_{s}(t-s)} =\frac{1}{2} \gamma_{0} \lambda \mathrm{e}^{-\lambda\left|t-s\right|}
\end{equation}
We define the function $\tilde{G}(t) = G(t) e^{-i h t}$, where $c_{1}(t)=G(t) c_{1}(0)$, which can show that it defined as the solution of the integro–differential equation,
\begin{equation}\label{diff_gtelda}
\partial_{t} \tilde{G}(t)=-\int_{0}^{t} d s \left\langle h^{-}(t) h^{+}(s)\right\rangle_{E} e^{i \omega_{s}(t-s)} e^{-i h(t-s)} \tilde{G}(s)
\end{equation}
with initial condition $\tilde{G}(0)=G(0)=1$. Generally, $G(t)$ can be solved to give the exact solution by Laplace transform. Indeed, substituting Eq. \eqref{General_f}  into Eq. \eqref{diff_gtelda} we obtain,
\begin{equation}\label{propagator_g}
G(t)= e^{-\lambda  t/2} \left(\frac{  \sinh \left(\frac{\lambda \chi }{2} t\right)}{\chi}+\cosh \left(\frac{\lambda \chi }{2} t \right)  \right)      
\end{equation}
Where $\chi = \sqrt{1-2 (\gamma_0/\lambda)}$.\\
In order to get a master equation in differential form with
a generator local in time, that is, the Time Convolutionless (TCL) master equation. We first give the exact time evolution mapping \cite{smirne2010nakajima}, which transforms the initial states into the states at time t
\begin{align}
\Phi(t): \rho(0) \rightarrow \rho(t)&= \operatorname{Tr}_E\left\lbrace\mid \psi(t)\rangle \langle\psi(t)\mid\right\rbrace \\&=\Phi(t) \rho(0), \: t\ge 0\notag
\end{align}
Then due to  Eq. \eqref{psi_t}, the density matrix $\rho(t)$ it is expressed as
\begin{equation}\label{rho_t}
\begin{split}
\rho (t)&=\operatorname{Tr}_{E}\{|\Psi(t)\rangle\langle\Psi(t)|\}\\
&=\left( \begin{array}{cc}{\rho }_{11}(t) & {\rho }_{10}(t)\\ {\rho }_{01}(t) & {\rho }_{00}(t)\end{array}\right)\\
&=\left(\begin{array}{cc}
|G(t)|^{2} \rho_{11}(0) & G(t) \rho_{10}(0) \\
G^{\star}(t) \rho_{01}(0) & \rho_{00}(0)+\left(1-|G(t)|^{2}\right) \rho_{11}(0)
\end{array}\right)\\
&=\left( \begin{array}{cc}|{c}_{1}(t{)|}^{2} & {c}_{0}^{\ast }{c}_{1}(t)\\ {c}_{0}{c}_{1}^{\ast }(t) & 1-|{c}_{1}(t{)|}^{2}\end{array}\right)
\end{split}
\end{equation}
where $ \rho_{i j}(t)=\langle i|\rho(t)| j\rangle$  for  i, j=0,1. We can construct the exact TCL equation, $\partial_{t} \rho(t)=$ $\mathcal{K}_{\mathrm{TCL}}(t) \rho(t)$, exploiting the  introduced relations \cite{smirne2010nakajima, Breuer}, we can introduce a time-local generator
\begin{equation}
\mathcal{K}_{\mathrm{TCL}}(t)=\dot{\Phi}(t) \Phi^{-1}(t)
\end{equation}
We can obtain an exact TCL master equation to second order in the interaction picture \cite{jing2018decoherence, vacchini2010exact, tong2010mechanism}
\begin{align}
\partial_{t} \rho(t)&=\mathcal{K}_{\mathrm{TCL}}(t) \rho(t)\\&=-\frac{i}{2} \varepsilon(t)\left[S_{+} S_{-}, \rho(t)\right]\notag\\&+\gamma(t) \left[ S_{-} \rho(t) S_{+}-\frac{1}{2}\lbrace S_{+} S_{-}, \rho(t)\rbrace\right]\notag
\end{align}
Consider the time derivative of $\rho(t)$
\begin{widetext}
\begin{equation}
{\partial }_{t}\rho (t)=\left( \begin{array}{cc} \Re(\frac{{\dot{c}}_{1}(t)}{{c}_{1}(t)})|{c}_{1}(t{)|}^{2} & (\frac{{\dot{c}}_{1}(t)}{{c}_{1}(t)}){c}_{0}^{\ast }{c}_{1}(t)\\ (\frac{{\dot{c}}_{1}^{\ast }(t)}{{c}_{1}^{\ast }(t)}){c}_{0}{c}_{1}^{\ast }(t) & -\Re(\frac{{\dot{c}}_{1}(t)}{{c}_{1}(t)})|{c}_{1}(t{)|}^{2}\end{array}\right)=\left(\begin{array}{cc}
\gamma(t)\left|c_{1}(t)\right|^{2} & \left(\frac{i}{2} \varepsilon (t)-\frac{1}{2} \gamma(t)\right) c_{0} c_{1}^{*}(t) \\
\left(-\frac{i}{2} \varepsilon(t)-\frac{1}{2} \gamma(t)\right) c_{0}^{*} c_{1}(t) & -\gamma(t)\left|c_{1}(t)\right|^{2}
\end{array}\right)
\end{equation}
\end{widetext}
The dynamics of the exact TCL master equation, parameterized by $\epsilon(t)$ plays the role of a time-dependent Lamb shift, which is induced by coupling to the noisy environments, and $\gamma(t)$ plays the role of a time-dependent decay rate(decoherence rate). In addition, the decay rate $\gamma(t)$ can have negative values, which means that the dynamics of the system exhibit a strong non-Markovian behavior~\cite{vacchini2010exact}. This parameters can be written as follows,
\begin{align}
\gamma(t)+i\varepsilon(t) &=-2\left[\frac{\dot{G}(t)}{G(t)}\right], \: \varepsilon(t)=-2 \Im\left[\frac{\dot{G}(t)}{G(t)}\right], \\& \gamma(t)=-2 \Re\left[\frac{\dot{G}(t)}{G(t)}\right]\notag
\end{align}
We use the fidelity~\cite{Jing2013, jing2018decoherence} $\mathcal{F}(t) = \sqrt{\langle\psi(0)|\rho(t)| \psi(0)\rangle}$ to measure the decoherence dynamics of the
central spin. Indeed, the coherence of spin qubits is highly affected by the nuclear spins of the host material and their hyperfine coupling to the electron spin. When the system is prepared in the initial state $\ket{\psi(0)}= \ket{1}$, the fidelity can be written as
\begin{equation}\label{fid}
\mathcal{F}(t)=\sqrt{\left|c_{1}(t) c_{1}(0)\right|^{2}}=|G(t)|=|\tilde{G}(t)|
\end{equation}
The fidelity illustrated in Fig. \eqref{Fidlity_general} and Fig. \eqref{Fidlity_gamma_0_t}, can be considered in two main regimes. In the weak coupling case, which reflects that the decoherence dynamics of the quantum system is Markovian, $\lambda > 2 \gamma_0$, one has $\chi\, \epsilon\, \mathbb{R}$ ($\chi = \sqrt{1-2 \gamma_0/\lambda}$)  so that $G(t)$ is
always positive. Indeed, when there is no more exchange between the qubit and his environment the coupling strength $\gamma_0 \rightarrow 0$ the fidelity $\lim_{\gamma_0\to 0} \mathcal{F}(t)\sim 1$. However, in the strong coupling case, $\lambda <2 \gamma_0$ one has $\chi\, \epsilon\, \mathrm{i}\mathbb{R}$, so that $G(t)$ oscillates between positive and negative values going through zero. This means the fidelity $\mathcal{F}(t)$ will then decay with an oscillating. Indeed in Fig. \eqref{Fidlity_general} by observing the contrast between the light blue and dark blue areas when $\lambda/\gamma_0<1$ , in another word the fidelity has a revival, which means the quantum information flow bounces from the spin bath back to the qubit system. This is the signature of non-Markovian behavior.

We consider the case which describes the central spin of MoS$_2$ QD qubit overlaps with about $\mathcal{N}$ nuclear spins and they interact via hyperfine interaction. This may lead to entanglement between the qubit and the nuclear bath and to back-action effects from the qubit to the nuclei and vice versa. In this situation, we assume that the hyperfine interaction strength $A_{k} \approx A/ \mathcal{N}$ and the nuclear Zeeman splitting $\omega_{k} = g_{I_{k}} \mu_{N} b_z$ satisfies a Gaussian distribution characterized by the mean value $\bar{\omega}$ and the parameter $\nu^2$ is referred to as the variance, where $\bar{\omega}$ and $\nu$ can be supposed to be in the same order of $\approx |A| / \sqrt{\mathcal{N}}$. Indeed, the electron Zeeman splitting $\omega_{s} \approx A=\mathcal{N} A_{k} \approx \sqrt{\mathcal{N}} \omega_{k}$ are much larger than  nuclear splitting $\omega_{k}$ which can be approximated as a continuous variable centering around the average value $\sim A / \sqrt{\mathcal{N}}$ following Gaussian distribution, 
$\mathcal{P}(\omega)=1/(\sqrt{2 \pi} \nu) e^{-\frac{(\omega-\bar{\omega})^{2}}{2 \nu^{2}}} $.  The correlation function can be expressed as follows
\begin{equation}
\left\langle h^{-}(t) h^{+}(s)\right\rangle_{E} e^{i \omega_{s}(t-s)} e^{-i h(t-s)}=\sum_{k=0}^{\mathcal{N}-1} (\frac{A_{k}}{2})^{2} e^{i \Omega_{k}(t-s)}
\end{equation}
where the effective detuning $\Omega_{k}=\omega_{s}-\omega_{k}+A_{k} / 2-h$, is usually understood as a measure of the memory capacity or non-Markovianity of the environment. Then, the effective correlation function of the spin bath can then be evaluated as
\begin{widetext}
\begin{equation}\label{f_case2}
\left\langle h^{-}(t) h^{+}(s)\right\rangle_{E} e^{i \omega_{s}(t-s)} e^{-i h(t-s)} \approx \frac{A^{2}}{4 N} e^{i (\omega_{s}+\frac{A}{2\mathcal{N}}-h)(t-s)} \int d \omega \mathcal{P}(\omega) e^{-i \omega(t-s)} 
\approx \frac{A^{2}}{4 N} e^{-\frac{\nu^{2}}{2}(t-s)^{2}+i (\omega_{s}+\frac{A}{2\mathcal{N}}-h)(t-s)}
\end{equation}
\end{widetext}
The fidelity $\mathcal{F}(t)=|\widetilde{G}(t)|$ can be numerically obtained by inserting Eq. \eqref{f_case2} into Eq. \eqref{diff_gtelda}.

To perform the numerical simulation. We have chosen
the  MoS$_2$ parameters as follows. The strength of the hyperfine coupling A has been estimated to be $A = 0.29\, \mu$eV. This estimate is based on an average over the hyperfine coupling constants for the two  nuclear isotopes $^{95}$Mo and $^{97}$Mo, weighted by their relative abundance, see Eq. \eqref{A_hf}. The naturally occurring isotopes carry spin with  $I_{(^{95}\text{Mo})}=I_{(^{97}\text{Mo})}=5/2$. In this model, we have the Overhauser’s field $h\thickapprox I_0 A \thickapprox \frac{5}{2} A$. Here we study the case of a localized electron Spin trapped in quantum dot of MoS$_2$ with radius $R = 26 \,nm$ interacting with $\mathcal{N}$ polarized nuclear spin environments via hyperfine interaction, where we employ the same time-convolutionless (TCL) method as some previous work  \cite{jing2018decoherence}.
\begin{figure}[h!]%
    \centering
\includegraphics[width=\linewidth]{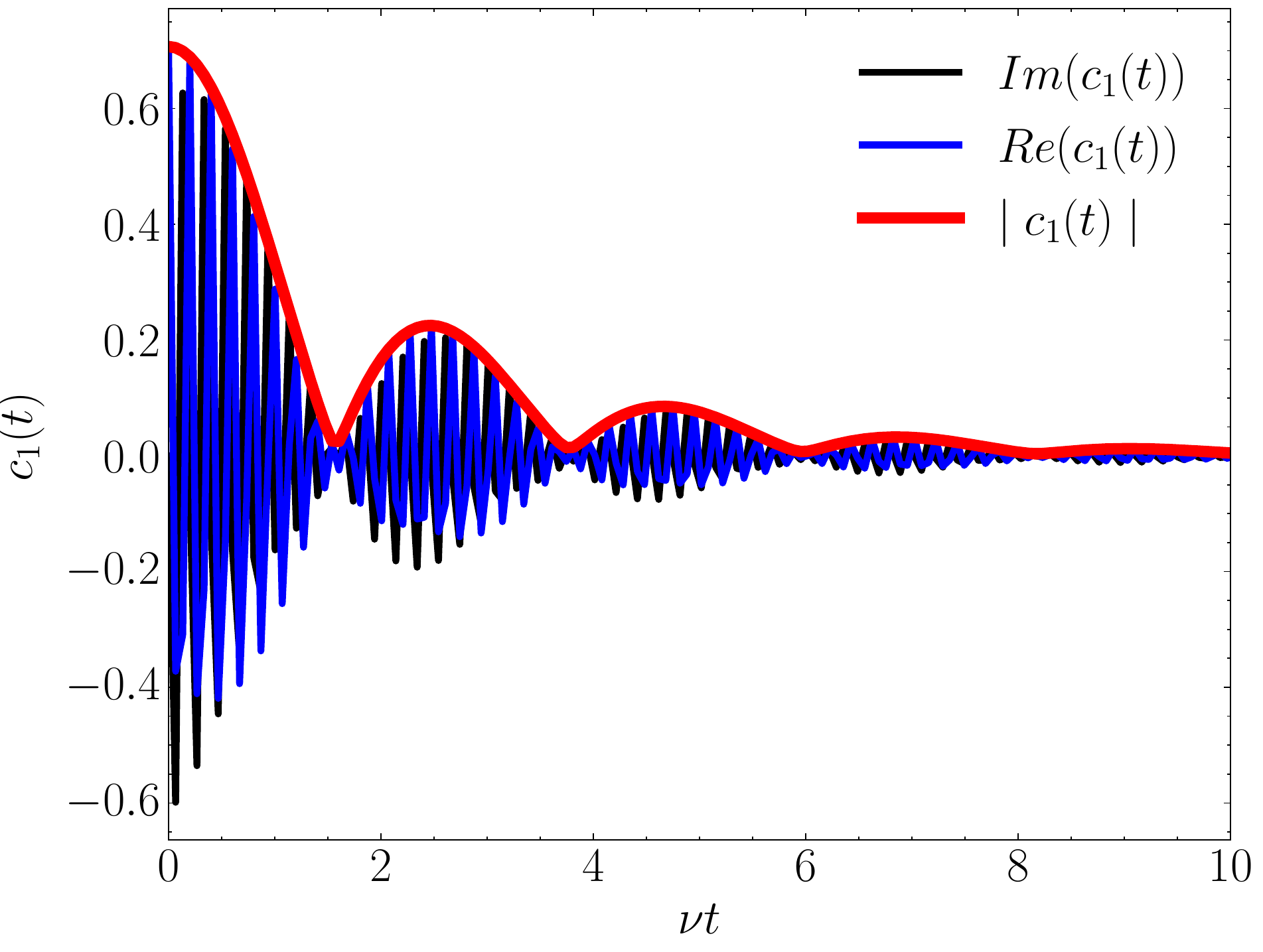}
    \caption{\label{fig:c0t_1}Time evolution of the amplitude $c_1(t)$  as a function of dimensionless time $\nu t$. Blue(black)lines: Real(imaginary) part of the amplitude $c_1(t)$. The initial state $\ket{\psi(0)}=\frac{1}{\sqrt{2}}(\mid 0\rangle+ \mid 1\rangle)$, $c_0=c_1(0) = 1/\sqrt{2}$. Here, we have $\mathcal{N} = 10^2$.}%
\end{figure}
\begin{figure}[h!]%
    \centering
\includegraphics[width=\linewidth]{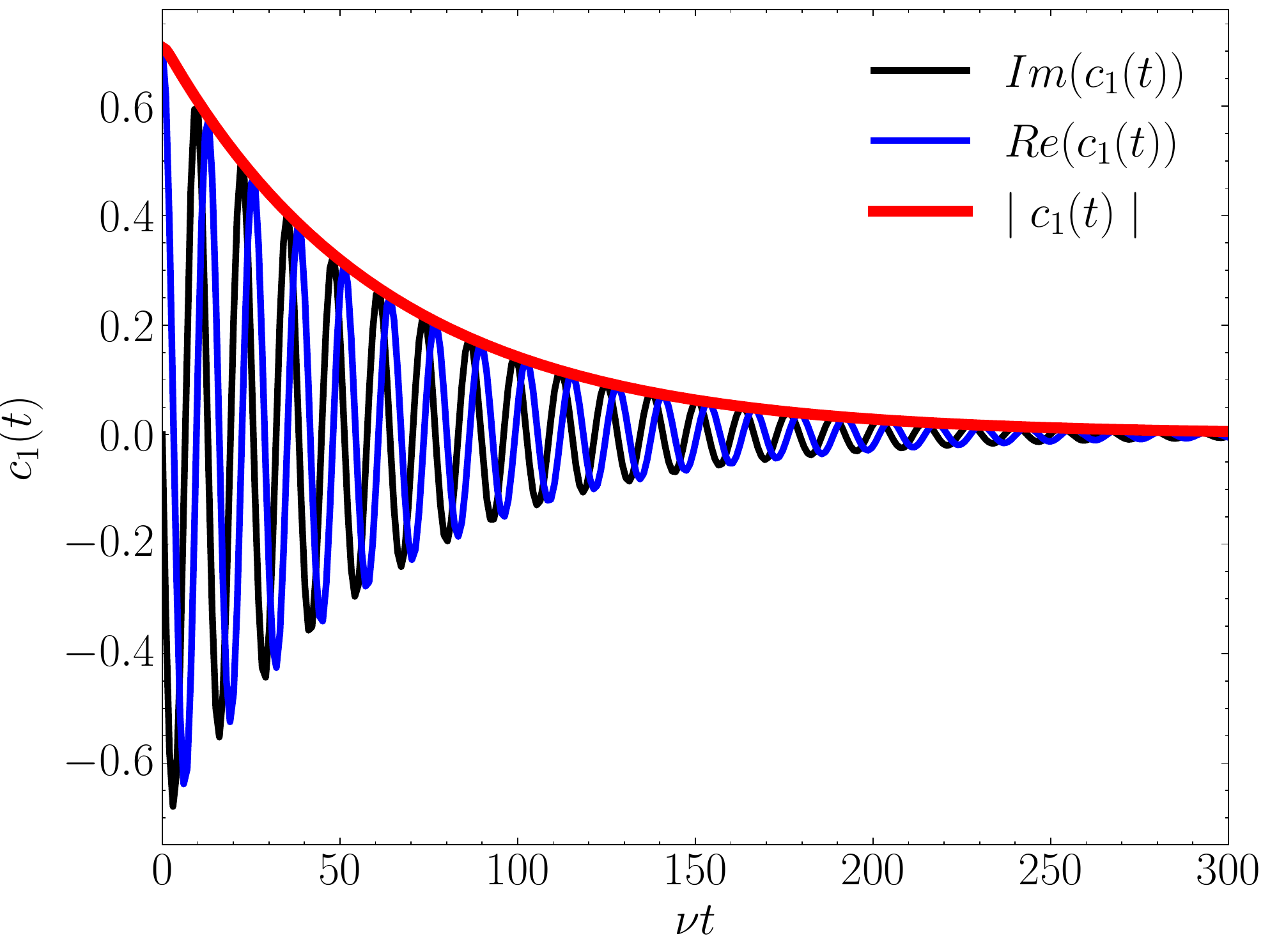}
    \caption{\label{fig:c0t_2}Time evolution of the amplitude $c_1(t)$  as a function of dimensionless time $\nu t$. Blue(black)lines: Real(imaginary) part of the amplitude $c_1(t)$. The initial state $\ket{\psi(0)}=\frac{1}{\sqrt{2}}(\mid 0\rangle+ \mid 1\rangle)$, $c_0=c_1(0) = 1/\sqrt{2}$.  Here, we have $\mathcal{N} = 10^4$.}%
\end{figure}

In Fig.~\eqref{fig:c0t_1} and Fig.~\eqref{fig:c0t_2} we show the dynamics of $c_1(t)$ as a function of dimensionless time $\nu t$ for different value of nuclear spins $\mathcal{N}$. Fig. \eqref{fig:c0t_1}, for $\mathcal{N} = 10^2$, shows the oscillation of the real and imaginary parts of the amplitude $c_1(t)$ decays non-exponentially and display a clear beating pattern. Such oscillation decreases gradually with the passage of time to some equilibrium value. $\mid c_1(t)\mid $ shows nonmonotonic oscillatory decay with zero coherence revivals, which occurs by the electron spin-flip  transition. Remarkably, the dynamics describes the initial oscillations of the $c_1(t)$ appear in the non-Markovian description of open quantum systems.  However,  for $\mathcal{N} = 10^4$ Fig. \eqref{fig:c0t_2} shows the oscillation, where the decay rate of amplitude $c_1(t)$  becomes much bigger compared to Fig.~\eqref{fig:c0t_1} which shown by the transition from nonmonotonic oscillatory decay to monotonic decay. Physically this means, that increasing the number of nuclear spins $\mathcal{N}$  can suppress the quantum fluctuations which induced by the nuclear dynamics and raise the qubit coherence.

\begin{figure}[htpb]
\subfigure[]{\includegraphics[width=\linewidth]{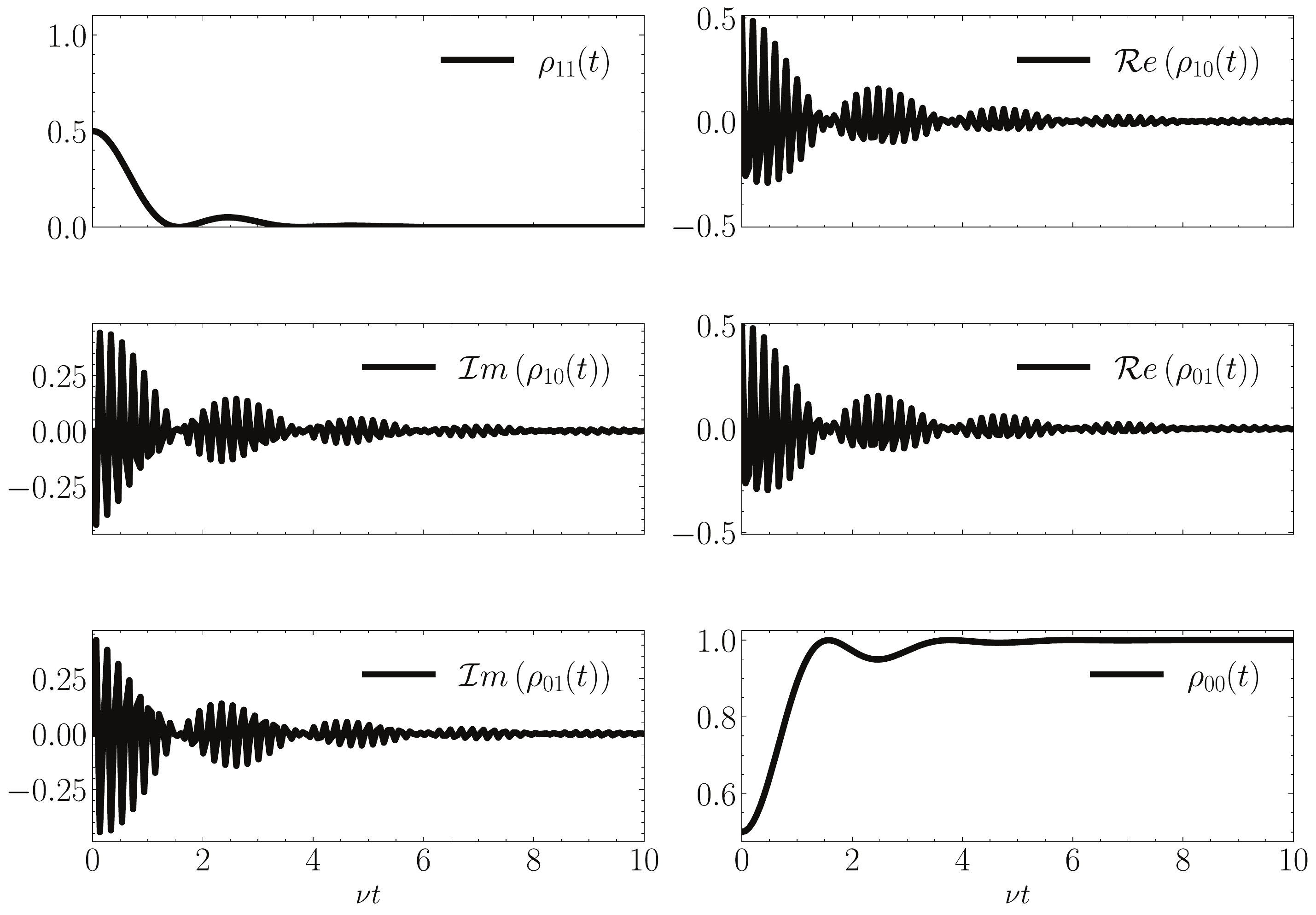}\label{ρ_matrix_element_t_case2_a} }
\subfigure[]{\includegraphics[width=\linewidth]{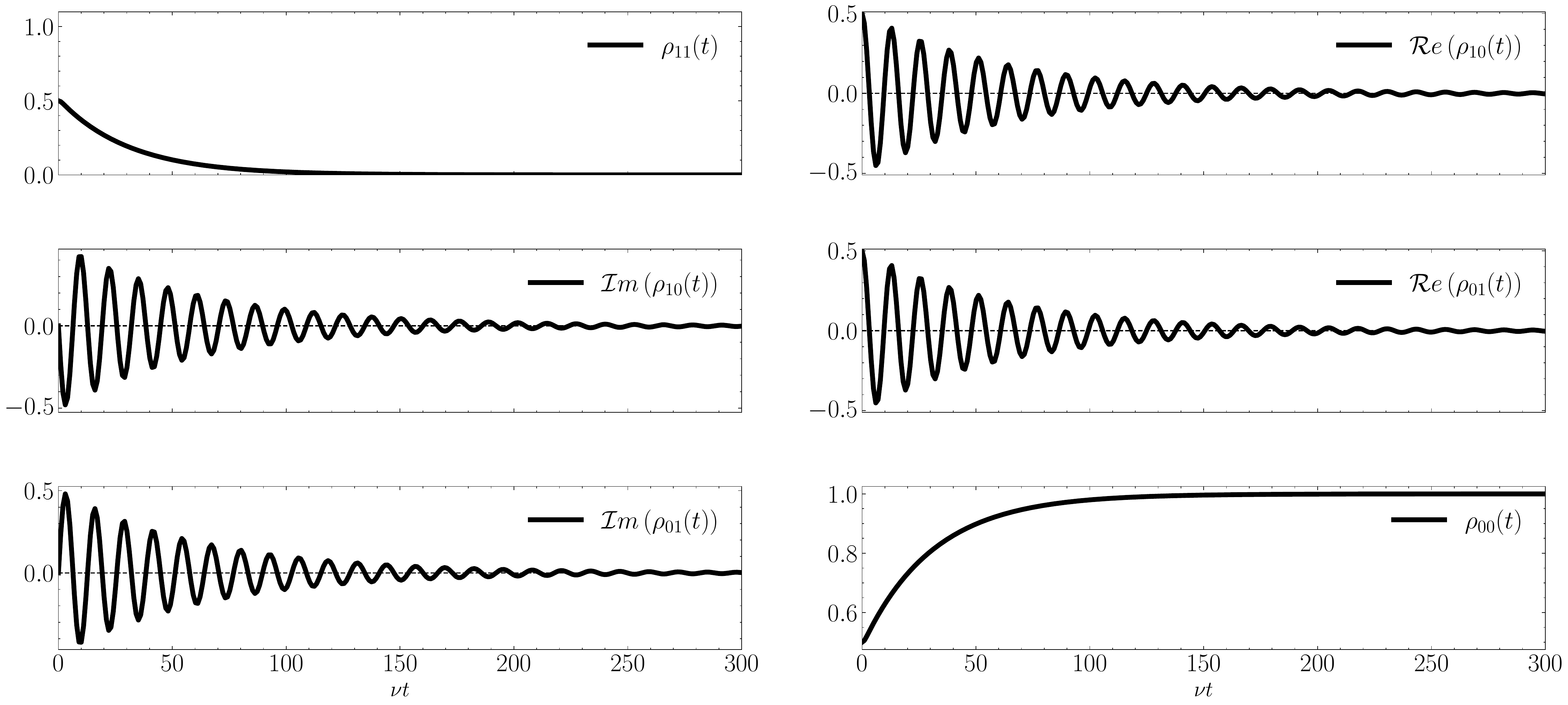}\label{ρ_matrix_element_t_case2_b} }
\caption{Time evolution of density matrix elements $\rho_{ij}$ as a function of dimensionless time $\nu t$. We give results from numerical Eq.~\eqref{rho_t}. The initial state $\ket{\psi(0)}=\frac{1}{\sqrt{2}}(\mid 0\rangle+ \mid 1\rangle)$, $c_0=c_1(0) = 1/\sqrt{2}$. (a)  Here, we have $\mathcal{N} = 10^2$. (b)  Here, we have $\mathcal{N} = 10^4$.}
\label{ρ_matrix_element_t_case2}
\end{figure}

To better understand the effect of the environment of nuclear spins by hyperfine interaction on the coherence of the MoS$_2$ spin-valley qubit system, we determine the temporal evolution of the density matrix $\rho(t)$. Fig.~$\eqref{ρ_matrix_element_t_case2}$ illustrate the evolution of populations (diagonal elements of the density matrix) and coherence (non-diagonal elements of the density matrix) as a function of dimensionless time $\nu t$. For non-diagonal elements shows in Fig.~\eqref{ρ_matrix_element_t_case2_a}, the oscillation decays non-exponentially and display a clear beating. These beating patterns clearly originate from the peaked nature of the environmental spectrum. The $\rho_{11}(t)$ shows nonmonotonic oscillatory decay with zero coherence revivals, which reflects that the decoherence dynamics of the quantum
system is non-Markovian. However, by increasing the number of nuclear spins $\mathcal{N}$ and examining the Fig.~$\eqref {ρ_matrix_element_t_case2_b} $, we can see that the non-diagonal elements exhibit a damped oscillatory behavior with the disappearance of the beat pattern. We can attribute this effect to the fact that when we increase $\mathcal{N}$, we go from a strong coupling regime between the qubit and the noisy environment described by the non-Markovian character to a weak coupling regime described by the Markovian character. 
\begin{figure}[h!]%
\centering
\includegraphics[width=\linewidth]{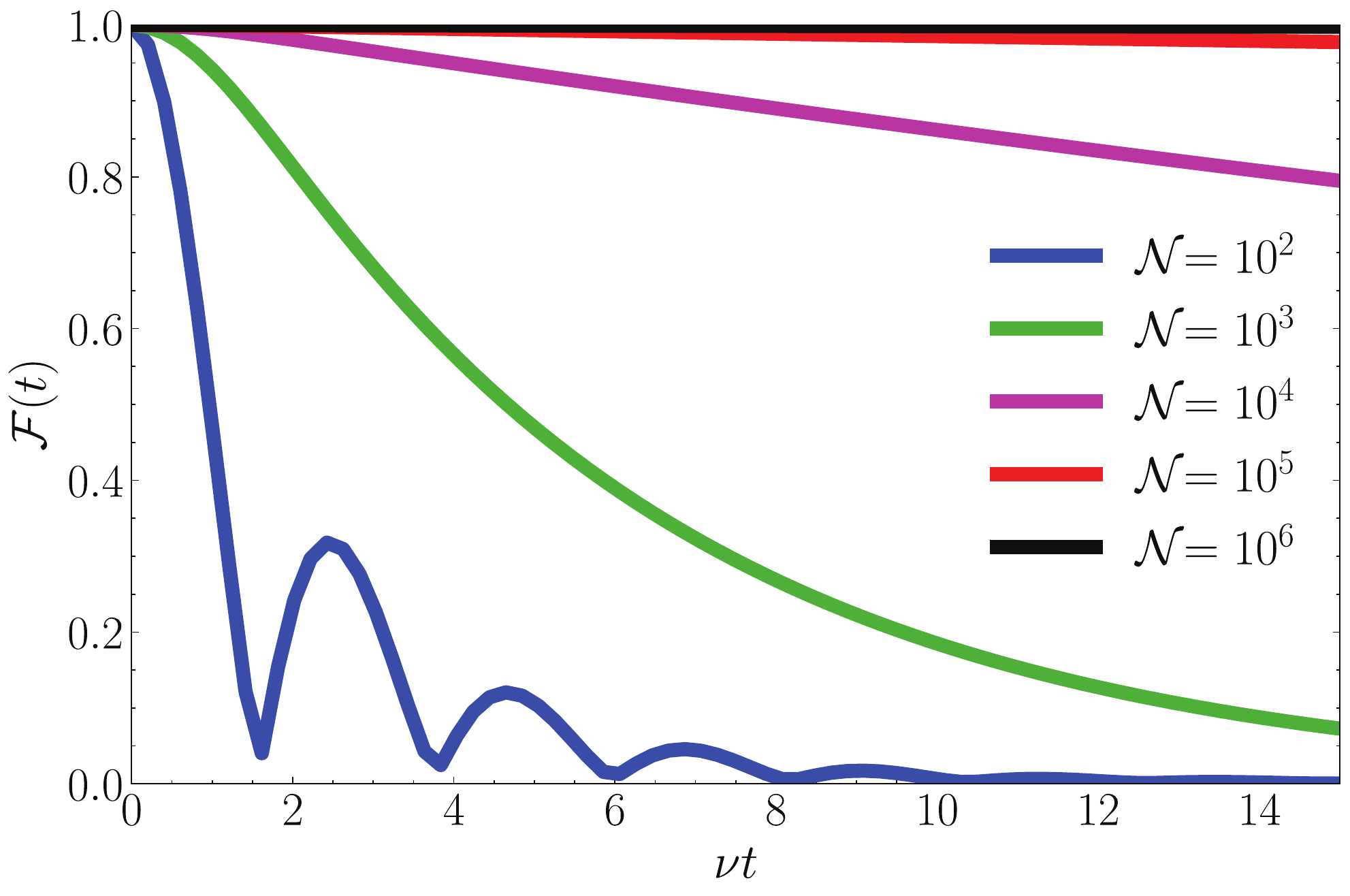}
\caption{ Time evolution of the fidelity, ${\mathcal F} (t)=|\tilde{G}(t)|$ as a function of dimensionless time $\nu t$ for different values of the number of nuclear spins within the bath $\mathcal{N}$. For a system is prepared in the initial state $\ket{\psi(0)}= \ket{1}$.}\label{Fidelity_t_case2}
\end{figure}

Fig.~\eqref{Fidelity_t_case2} show the fidelity $\mathcal{F}(t)$ as a function of dimensionless time $\nu t$ induced by the non-Markovian hyperfine interaction for different values of the number of nuclear spins $\mathcal{N}$ within the bath. The value of $\mathcal{F}(t)$ exactly reflects the decoherence of the central spin. The closer the value of $\mathcal{F}(t)$ to 1, the smaller the difference between the current state and the initial state of the central spin. As shown in Fig.~\eqref{Fidelity_t_case2} we can clearly see that when $\mathcal{N} = 10^2$, the fidelity shows nonmonotonic oscillatory decay with zero coherence revivals, meaning that the exchange of the quantum information and energy between the system and bath spins having a noticeable or major effect, the quantum information flow bounces from the spin bath back to the system. In contrast, when $\mathcal{N}$ increases the decay rate of the central spin becomes gradually smaller which shown by the transition from nonmonotonic oscillatory decay to monotonic decay. In addition, the coherence of the central spin remains robust for large $\mathcal{N}$, these results indicate that the fidelity improves as the effect of environmental memory increases. The spin-bath with $\mathcal{N}$ up to $10^6$ works as it's providing a natural protection for central spin coherence. The varying of the nuclear spins $\mathcal{N}$ value depends on QD radius R as shown in Sec.~\eqref{ME}, which mean by increasing R we increase $\mathcal{N}$. Knowing also, that choosing R is a prerequisite for realizing the MoS$_2$ spin-valley qubit, see Sec.~\eqref{Qubit_candidat}.

\section{Conclusion and outlook}\label{conclusion} % [Averge[3 Paragraphe]]
In this paper,  we considered an approach for the description of open quantum systems. Thus, we have proposed an exact master equation for a central spin coupled to a spin bath by hyperfine interaction, indeed this approach is often used to model noise in solid-state qubits. In particular, we study the time-convolutionless master equation for the full polarized environment bath. Although our master equation is obtained based on the full polarization assumption and restricted to the second order, yet it still provides a convenient way to emphasize and capture the non-Markovian feature of the nuclear spin bath.\\
By investigating a ML-MoS$_{2}$ Spin-valley QD, with an external magnetic field applied perpendicular, which can then be used to tune the energy splitting between these two states.\\
We analyze the decoherence dynamics from induced noise determined by the correlation function corresponding to this spin bath model, uniform hyperfine interaction strength and Gaussian distribution in terms of bath-spin frequency, respectively. Described by this noise, the effect of the spin bath on the central spin gives rise to a reduced dynamics. Furthermore, We have found that the Overhauser's field in QD system may help to restrict the decoherence process of the central electron spin, which can regain its coherence and retain its initial state under an environment with a larger number of bath spins.  An obvious extension of this work is to use the fidelity to explicitly show the signature of non-Markovian behavior. As a consequence of this, the environmental non-Markovian feature can increase the coherence in the single qubit dynamics.\\
This model is qualitatively valid for other systems fulfilling the requirements of this later, where the most important demands are a Gaussian-like envelope function, slow dynamics of the nuclear bath, and a sufficiently large Zeeman-splitting with respect to the HI energy scale.
Our results are helpful for further understanding the non-Markovian qubit dynamics in the presence of non-equilibrium environments.

\bibliographystyle{unsrt}
\bibliography{apssamp}

\providecommand{\noopsort}[1]{}\providecommand{\singleletter}[1]{#1}%
\begin{thebibliography}{10}

\bibitem{lee2017valley}
Jieun Lee, Zefang Wang, Hongchao Xie, Kin~Fai Mak, and Jie Shan.
\newblock Valley magnetoelectricity in single-layer mos 2.
\newblock {\em Nature materials}, 16(9):887--891, 2017.

\bibitem{Valleytronics}
John~R. Schaibley, Hongyi Yu, Genevieve Clark, Pasqual Rivera, Jason~S. Ross,
  Kyle~L. Seyler, Wang Yao, and Xiaodong Xu.
\newblock Valleytronics in 2d materials.
\newblock {\em Nature Reviews. Materials}, 1(11), 8 2016.

\bibitem{Korm_nyos_2014}
Andor Kormányos, Viktor Zólyomi, Neil~D. Drummond, and Guido Burkard.
\newblock Spin-orbit coupling, quantum dots, and qubits in monolayer transition
  metal dichalcogenides.
\newblock {\em Physical Review X}, 4(1), Mar (2014).

\bibitem{Kormanyos_2018}
Alessandro David, Guido Burkard, and Andor Kormányos.
\newblock Effective theory of monolayer tmdc double quantum dots.
\newblock {\em 2D Materials}, 5, 02 2018.

\bibitem{Impurity_2018}
G{\'a}bor Sz{\'e}chenyi, Luca Chirolli, and Andras P{\'a}lyi.
\newblock Impurity-assisted electric control of spin-valley qubits in monolayer
  mos2.
\newblock {\em 2D Materials}, 5(3):035004, 2018.

\bibitem{PhysRevB}
Yue Wu, Qingjun Tong, Gui-Bin Liu, Hongyi Yu, and Wang Yao.
\newblock Spin-valley qubit in nanostructures of monolayer semiconductors:
  Optical control and hyperfine interaction.
\newblock {\em Phys. Rev. B}, 93:045313, Jan 2016.

\bibitem{pearce2017electron}
Alexander~J Pearce and Guido Burkard.
\newblock Electron spin relaxation in a transition-metal dichalcogenide quantum
  dot.
\newblock {\em 2D Materials}, 4(2):025114, 2017.

\bibitem{pawlowski2018valley}
Jaroslaw Paw{\l}owski, Dariusz {\.Z}ebrowski, and Stanis{\l}aw Bednarek.
\newblock Valley qubit in a gated mos 2 monolayer quantum dot.
\newblock {\em Physical Review B}, 97(15):155412, 2018.

\bibitem{Brooks_2017}
Matthew Brooks and Guido Burkard.
\newblock Spin-degenerate regimes for single quantum dots in transition metal
  dichalcogenide monolayers.
\newblock {\em Physical Review B}, 95(24), Jun (2017).

\bibitem{coish2004hyperfine}
WA~Coish and Daniel Loss.
\newblock Hyperfine interaction in a quantum dot: Non-markovian electron spin
  dynamics.
\newblock {\em Physical Review B}, 70(19):195340, (2004).

\bibitem{fuchs2012spin}
Moritz Fuchs, Valentin Rychkov, and Bj{\"o}rn Trauzettel.
\newblock Spin decoherence in graphene quantum dots due to hyperfine
  interaction.
\newblock {\em Physical Review B}, 86(8):085301, (2012).

\bibitem{wu2016spin}
Yue Wu, Qingjun Tong, Gui-Bin Liu, Hongyi Yu, and Wang Yao.
\newblock Spin-valley qubit in nanostructures of monolayer semiconductors:
  Optical control and hyperfine interaction.
\newblock {\em Physical Review B}, 93(4):045313, 2016.

\bibitem{RevModPhys.85.79}
Bernhard Urbaszek, Xavier Marie, Thierry Amand, Olivier Krebs, Paul Voisin,
  Patrick Maletinsky, Alexander H\"ogele, and Atac Imamoglu.
\newblock Nuclear spin physics in quantum dots: An optical investigation.
\newblock {\em Rev. Mod. Phys.}, 85:79--133, Jan 2013.

\bibitem{PhysRevA.81.042103}
Bassano Vacchini and Heinz-Peter Breuer.
\newblock Exact master equations for the non-markovian decay of a qubit.
\newblock {\em Phys. Rev. A}, 81:042103, Apr 2010.

\bibitem{Shenarticle2014}
H.~Shen, Mingzuo Qin, Xiaoming Xiu, and X.~Yi.
\newblock Exact non-markovian master equation for a driven damped two-level
  system.
\newblock {\em Physical Review A}, 89, 06 2014.

\bibitem{smirne2010nakajima}
Andrea Smirne and Bassano Vacchini.
\newblock Nakajima-zwanzig versus time-convolutionless master equation for the
  non-markovian dynamics of a two-level system.
\newblock {\em Physical Review A}, 82(2):022110, (2010).

\bibitem{BLAHA1990399}
P.~Blaha, K.~Schwarz, P.~Sorantin, and S.B. Trickey.
\newblock Full-potential, linearized augmented plane wave programs for
  crystalline systems.
\newblock {\em Computer Physics Communications}, 59(2):399--415, 1990.

\bibitem{blaha2001wien2k}
Peter Blaha, Karlheinz Schwarz, Georg~KH Madsen, Dieter Kvasnicka, Joachim
  Luitz, et~al.
\newblock wien2k.
\newblock {\em An augmented plane wave+ local orbitals program for calculating
  crystal properties}, 60, 2001.

\bibitem{perdew1992atoms}
John~P Perdew, John~A Chevary, Sy~H Vosko, Koblar~A Jackson, Mark~R Pederson,
  Dig~J Singh, and Carlos Fiolhais.
\newblock Atoms, molecules, solids, and surfaces: Applications of the
  generalized gradient approximation for exchange and correlation.
\newblock {\em Physical review B}, 46(11):6671, 1992.

\bibitem{Coish}
W.~Coish, Jan Fischer, and Daniel Loss.
\newblock Exponential decay in a spin bath.
\newblock {\em Physical Review B}, 77, 11 2007.

\bibitem{Sarmaarticle}
Lukasz Cywinski, Wayne Witzel, and Sankar Das~Sarma.
\newblock Pure quantum dephasing of a solid state electron spin qubit in a
  large nuclear spin bath coupled by long-range hyperfine-mediated
  interactions.
\newblock {\em Physical Review B}, 79, 03 2009.

\bibitem{urbaszek2013nuclear}
Bernhard Urbaszek, Xavier Marie, Thierry Amand, Olivier Krebs, Paul Voisin,
  Patrick Maletinsky, Alexander H{\"o}gele, and Atac Imamoglu.
\newblock Nuclear spin physics in quantum dots: An optical investigation.
\newblock {\em Reviews of Modern Physics}, 85(1):79, 2013.

\bibitem{Taylor_2003}
J.~M. Taylor, C.~M. Marcus, and M.~D. Lukin.
\newblock Long-lived memory for mesoscopic quantum bits.
\newblock {\em Physical Review Letters}, 90(20), May (2003).

\bibitem{schliemann2003electron}
John Schliemann, Alexander Khaetskii, and Daniel Loss.
\newblock Electron spin dynamics in quantum dots and related nanostructures due
  to hyperfine interaction with nuclei.
\newblock {\em Journal of Physics: Condensed Matter}, 15(50):R1809, 2003.

\bibitem{ye2019spin}
Meng Ye, Hosung Seo, and Giulia Galli.
\newblock Spin coherence in two-dimensional materials.
\newblock {\em npj Computational Materials}, 5(1):1--6, (2019).

\bibitem{Coish2009}
W.~Coish and Jonathan Baugh.
\newblock Nuclear spins in nanostructures.
\newblock {\em physica status solidi (b)}, 246:2203, 10 2009.

\bibitem{avdeev2019hyperfine}
Ivan~D Avdeev and Dmitry~S Smirnov.
\newblock Hyperfine interaction in atomically thin transition metal
  dichalcogenides.
\newblock {\em Nanoscale Advances}, 1(7):2624--2632, (2019).

\bibitem{jing2018decoherence}
Jun Jing and Lian-Ao Wu.
\newblock Decoherence and control of a qubit in spin baths: an exact master
  equation study.
\newblock {\em Scientific reports}, 8(1):1--10, (2018).

\bibitem{Breuer}
Francesco~Petruccione Heinz-Peter~Breuer.
\newblock {\em The theory of open quantum systems}.
\newblock Oxford University Press, (2002).

\bibitem{vacchini2010exact}
Bassano Vacchini and Heinz-Peter Breuer.
\newblock Exact master equations for the non-markovian decay of a qubit.
\newblock {\em Physical Review A}, 81(4):042103, 2010.

\bibitem{tong2010mechanism}
Qing-Jun Tong, Jun-Hong An, Hong-Gang Luo, and CH~Oh.
\newblock Mechanism of entanglement preservation.
\newblock {\em Physical Review A}, 81(5):052330, 2010.

\bibitem{Jing2013}
Jun Jing, Lian-Ao Wu, Marcelo Sarandy, and J.~Muga.
\newblock Inverse engineering control in open quantum systems.
\newblock {\em Physical Review A}, 88:053422, 11 2013.

\end{thebibliography}

\end{document}